\documentclass[  superscriptaddress, amsmath,amssymb, aps,twocolumn,longbibliography]{revtex4-1}
\usepackage{graphicx}
\usepackage{dcolumn}
\usepackage{amsthm}
\usepackage{bm}
\newcommand{\ket}[1]{|#1\rangle}

\pdfoutput=1
\usepackage{graphicx,graphics,epsfig,subfigure,times,bm,bbm,amssymb,amsmath,amsthm,mathrsfs,MnSymbol}
\usepackage{gensymb}
\usepackage{amsfonts}
\usepackage[matrix,frame,arrow]{xypic}
\usepackage[pdfstartview=FitH]{hyperref}
\hypersetup{
    colorlinks=true,       
    linkcolor=red,          
    citecolor=magenta,        
    filecolor=magenta,      
    urlcolor=cyan,           
    runcolor=cyan}
\usepackage{epstopdf}
\usepackage[pdftex]{color}
\usepackage{braket}
\usepackage{enumerate}
\usepackage[normalem]{ulem}
\usepackage[usenames,dvipsnames]{xcolor}
\usepackage{multirow}
\usepackage{mathtools}

\definecolor{orange}{rgb}{1,0.5,0}

\newcommand{\ignore}[1]{}
\usepackage{geometry}

\ignore{
\documentclass[eprintnumbers,amsmath,amssymb,onecolumn,a4paper,caption ]{article}
\usepackage{amsfonts}
\usepackage{amssymb}
\usepackage{mathrsfs}
\usepackage{mathbbold}
\usepackage{bbm}
\usepackage{mathrsfs}
\usepackage{dcolumn}
\usepackage{bm}
\usepackage{times,epsfig,amssymb,amsmath}
\usepackage{float}
\usepackage{subfigure}
\usepackage{geometry}\geometry{left=2.5cm,right=2.5cm,top=3cm,bottom=3cm}

\usepackage{color}

}

\begin{document}

\title{Stark many-body localization transitions in superconducting circuits}

\author{Yong-Yi Wang}
\affiliation{Institute of Physics, Chinese Academy of Sciences, Beijing 100190, China}
\affiliation{School of Physical Sciences, University of Chinese Academy of Sciences, Beijing 100190, China}

\author{Zheng-Hang Sun}
\email{zhsun@iphy.ac.cn}
\affiliation{Institute of Physics, Chinese Academy of Sciences, Beijing 100190, China}
\affiliation{School of Physical Sciences, University of Chinese Academy of Sciences, Beijing 100190, China}

\author{Heng Fan}
\email{hfan@iphy.ac.cn}
\affiliation{Institute of Physics, Chinese Academy of Sciences, Beijing 100190, China}
\affiliation{School of Physical Sciences, University of Chinese Academy of Sciences, Beijing 100190, China}
\affiliation{Songshan Lake  Materials Laboratory, Dongguan 523808, Guangdong, China}
\affiliation{CAS Center of Excellence for Topological Quantum Computation, University of Chinese Academy of Sciences,
Beijing 100190, China}

\begin{abstract}
\noindent Recent numerical and experimental works have revealed a disorder-free many-body localization (MBL) in an interacting system subjecting to a linear potential, known as the Stark MBL. The conventional MBL, induced by disorder, has been widely studied by using quantum simulations based on superconducting circuits. Here, we consider the Stark MBL in two types of superconducting circuits, i.e., the 1D array of superconducting qubits, and the circuit where non-local interactions between qubits are mediated by a resonator bus. We calculate the entanglement entropy and participate entropy of the highly-excited eigenstates, and obtain the lower bound of the critical linear potential $\gamma_{c}$, using the finite-size scaling collapse. Moreover, we study the non-equilibrium properties of the Stark MBL. In particular, we observe an anomalous relaxation of the imbalance, dominated by the power-law decay $t^{-\xi}$. The exponent $\xi$ satisfies $\xi\propto|\gamma-\gamma_{c}|^{\nu}$ when $\gamma<\gamma_{c}$, and vanishes for $\gamma\geq \gamma_{c}$, which can be employed to estimate the $\gamma_{c}$. Our work indicates that superconducting circuits are a promising platform for investigating the critical properties of the Stark MBL transition.
\end{abstract}
\pacs{Valid PACS appear here}
\maketitle

\section{Introduction}
A programmable superconducting circuit can be employed
in the demonstration of quantum supremacy~\cite{qs1,qs2,qs3}, the quantum
computation of electronic structure~\cite{qchemstry}, and the generation of
multipartite entangled states~\cite{GHZ1,GHZ2,GHZ3}. Superconducting circuits are
also an excellent platform for performing analog quantum
simulations~\cite{AQS1,AQS2,AQS3,AQS4,AQS_a1,AQS5,AQS6,AQS7,AQS_a2}, exploring quantum many-body systems out of
equilibrium~\cite{non_eq_sum1}.

Many-body localization (MBL) is an active subject in the field of
quantum simulation~\cite{AQS2,AQS4,AQS_a1,AQS5,cold_atom1,cold_atom2,cold_atom3,cold_atom4,cold_atom5,tapped_ion1,tapped_ion2}. MBL describes a
non-ergodic phase in interacting quantum systems, i.e., the
many-body localized phase, and is a counterexample
of quantum thermalization~\cite{review_mbl1,review_mbl2,review_mbl3,review_mbl4}. MBL can occur when
an interacting system subjects to sufficiently strong disorder,
including random and quasiperiodic fields~\cite{MBL_num1,MBL_num_a1}.
With the increase of disorder strength, there is an eigenstate phase transition from the ergodic to many-body localized phase.
For highly excited eigenstates of a disordered interacting system, the
ergodicity and MBL can be identified by
the volume and area law of the entanglement entropy (EE)~\cite{MBL_num1,MBL_num_a1,MBL_num2,MBL_num3,MBL_num4,MBL_num5,MBL_num_a2}, respectively,
as well as the scaling analysis of the participation entropy
(PE)~\cite{MBL_num5,MBL_num6}. More importantly, under the quench dynamics, MBL can be characterized by the slow relaxation of
imbalance and PE, retaining the memory of initial states~\cite{MBL_num_a2,MBL_num_dy1,MBL_num_dy2,MBL_num_dy3,MBL_num_dy4},
and the logarithmic growth of EE~\cite{MBL_num_dy5,MBL_num_dy6,MBL_num_dy7}. The characteristic non-equilibrium
properties of MBL pave the way to
studying the phenomenon using analog quantum simulations,
where out-of-equilibrium dynamics is naturally accessible.
MBL has been observed in the one-dimensional
(1D) array of superconducting qubits~\cite{AQS2,AQS5} and the
superconducting circuit with all-to-all connectivity~\cite{AQS4,AQS_a1}.

Recently, a number of attentions have been paid to explore
the many-body localization without disorder. A prominent
example is the Wannier-Stark MBL, achieved by a
linear field beyond a critical value~\cite{stark1,stark2,stark3,stark4,stark_a1}. The logarithmic growth
of EE and the finite stationary value of imbalance at long time, as non-equilibrium
characteristics of the Stark MBL, are
numerically revealed~\cite{stark1,stark2,jakub_add1}. Other experimental feasible probes of
the Stark MBL, including the dynamics of quantum
mutual information and the spin echo, are also explored~\cite{stark5}.
Besides aforementioned progress, however, critical
properties of the Stark MBL
transition are not well understood. How the critical value of
the linear field is extracted from its quench dynamics
remains an open question. Additionally, it is unclear whether
the critical value obtained from temporal evolution is
consistent with that yielded by the volume-to-area law
transition of the EE of highly excited eigenstates.

For the MBL due to disorder, the critical
properties of the ergodicity-MBL transition has been numerically
studied by employing the exact diagonalization (ED) for spin-1/2 chains with the size up to
24~\cite{MBL_num2,MBL_num6}, and more challenging Bose-Hubbard models with the
size up to 14, in cooperation with scaling collapse~\cite{stark_a1,BH_MBL1,BH_MBL2,BH_MBL3,BH_MBL4}.
Based on the increase of system size~\cite{qs1,AQS_a2}, superconducting circuits are potential to simulated
many-body systems in a regime not accessible with exact simulations on classical computers.
Thus, the finite-size effect, as a major obstacle of the ED calculation, can be overcome
by the quantum simulation using large-scale superconducting circuits.


In this work, we focus on two types of superconducting
circuits. The first is a 1D array of superconducting qubits,
described by the 1D Bose-Hubbard model. The signs of MBL are distinct when the qubit anharmonicity (in the
unit of the hopping interaction) is around 4~\cite{BH_MBL1}, which can be
realized in a device with coupler, enabling hopping
interactions to be tunable~\cite{AQS2}. Nevertheless, when the
anharmonicity is sufficiently large, the Bose-Hubbard model
is approximate to the hard-core limit, equivalent to the 1D
$XX$ spin model~\cite{BH_add}, as an noninteracting system~\cite{free_XY}. Consequently,
with large qubit anharmonicity, a 1D array of superconducting
qubits, subjecting to disorder, exhibits a marginally Anderson localization~\cite{anderson}.

For the superconducting qubit with large anharmonicity, the non-local
hopping interactions between qubits can render a nonintegrable
Hamiltonian, where thermalization and many-body
localization can be observed~\cite{AQS4,AQS_a1}. Thus, we also focus on
the superconducting circuit with all-to-all connectivity,
enabled by a resonator bus. Taking the hard-core limit into consideration,
the superconducting circuit can be modeled by a 1D $XX$ spin chain,
accompanying additional fully connected $XX$-type couplings.


We then calculate the EE and PE of the highly excited
eigenstates in the two types of superconducting circuits
using ED. By performing finite-size critical scaling collapse,
we obtain the critical linear field of the Stark
MBL transition. We also study the quench
dynamics of EE, PE, and imbalance. In particular, we
observe a power-law decay of the imbalance at
intermediate linear field, and further extract the critical
linear field from the fitted decay rate of the imbalance.

\section{The Stark many-body localization transition in a 1D array of superconducting
	qubits}

\subsection{Model}
\begin{figure}
	\centering
	\includegraphics[width=1\linewidth]{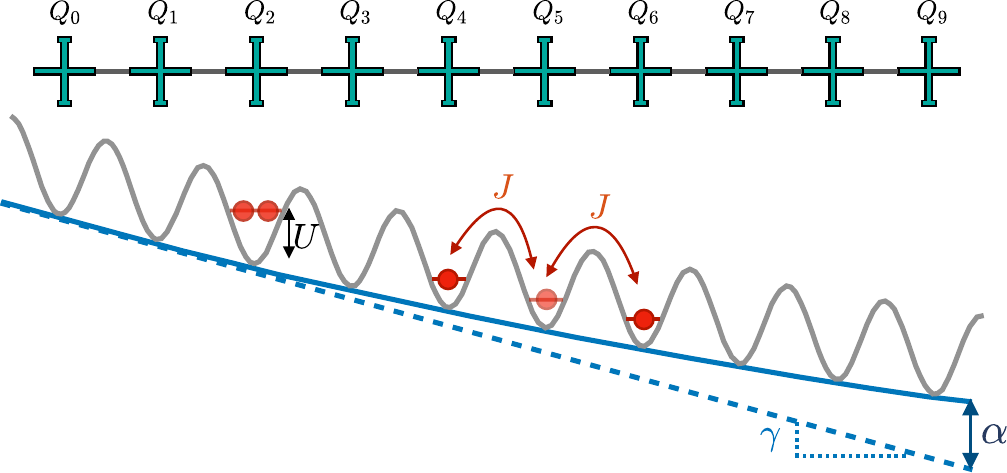}\\
	\caption{Top panel: A schematic illustration of a 1D array of 10 coupled superconducting qubits. Bottom panel: A sketch of the 1D Bose-Hubbard model subject to a linear potential with curvature, describing the superconducting circuit in the top panel. }\label{fig1}
\end{figure}

\begin{figure}
	\centering
	\includegraphics[width=1\linewidth]{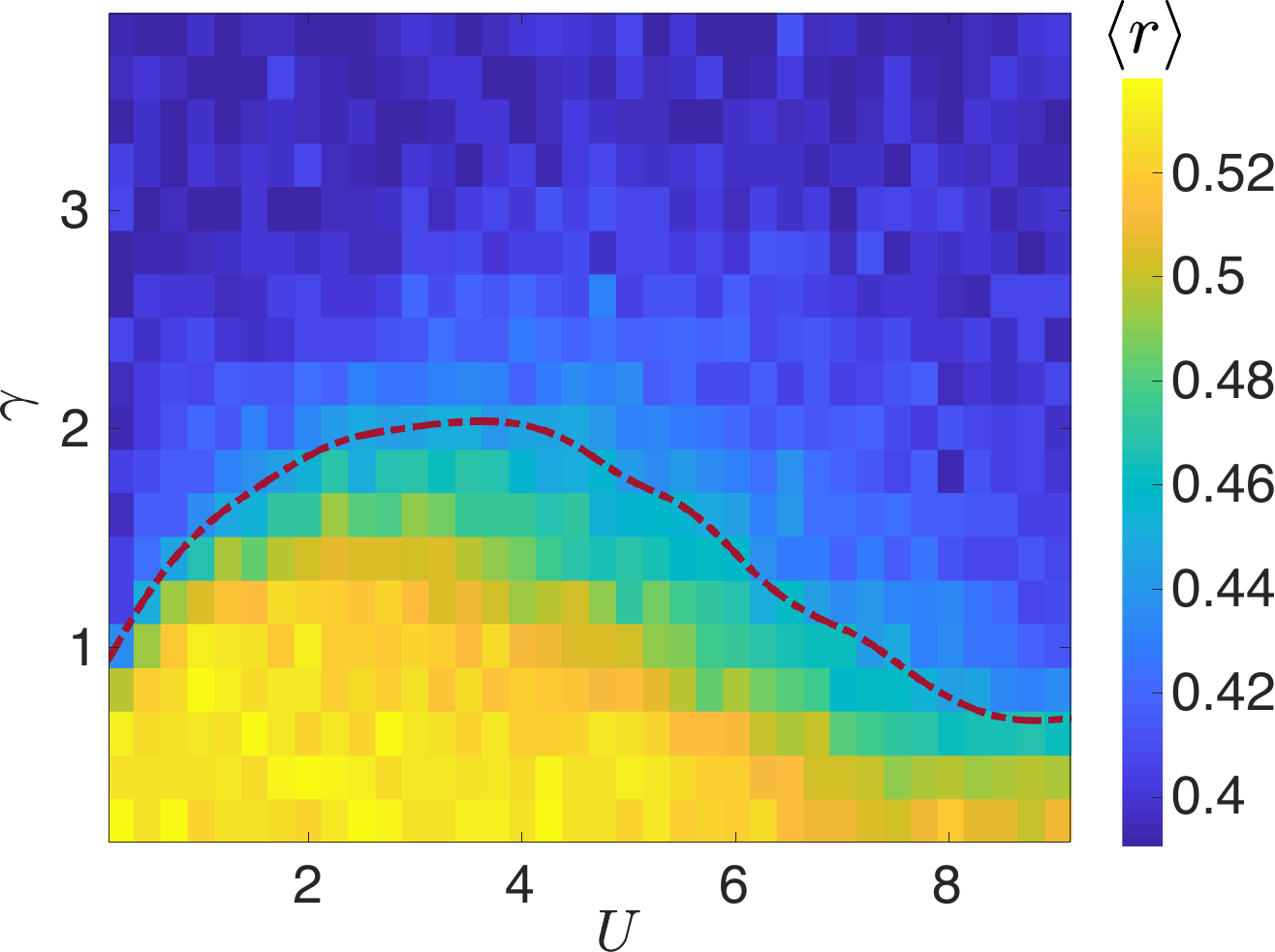}\\
	\caption{The ratio of consecutive level spacings averaged over a chosen energy window, i.e., $\langle r \rangle$, in the Hamiltonian (\ref{H_bh}) with $J=1$, $L=12$ and $\alpha=2$, as a function of $U$ and $\gamma$. The red dashed line is the contour for $\langle r\rangle\simeq 0.44$. }\label{fig2}
\end{figure}

\begin{figure*}
	\centering
	\includegraphics[width=0.9\linewidth]{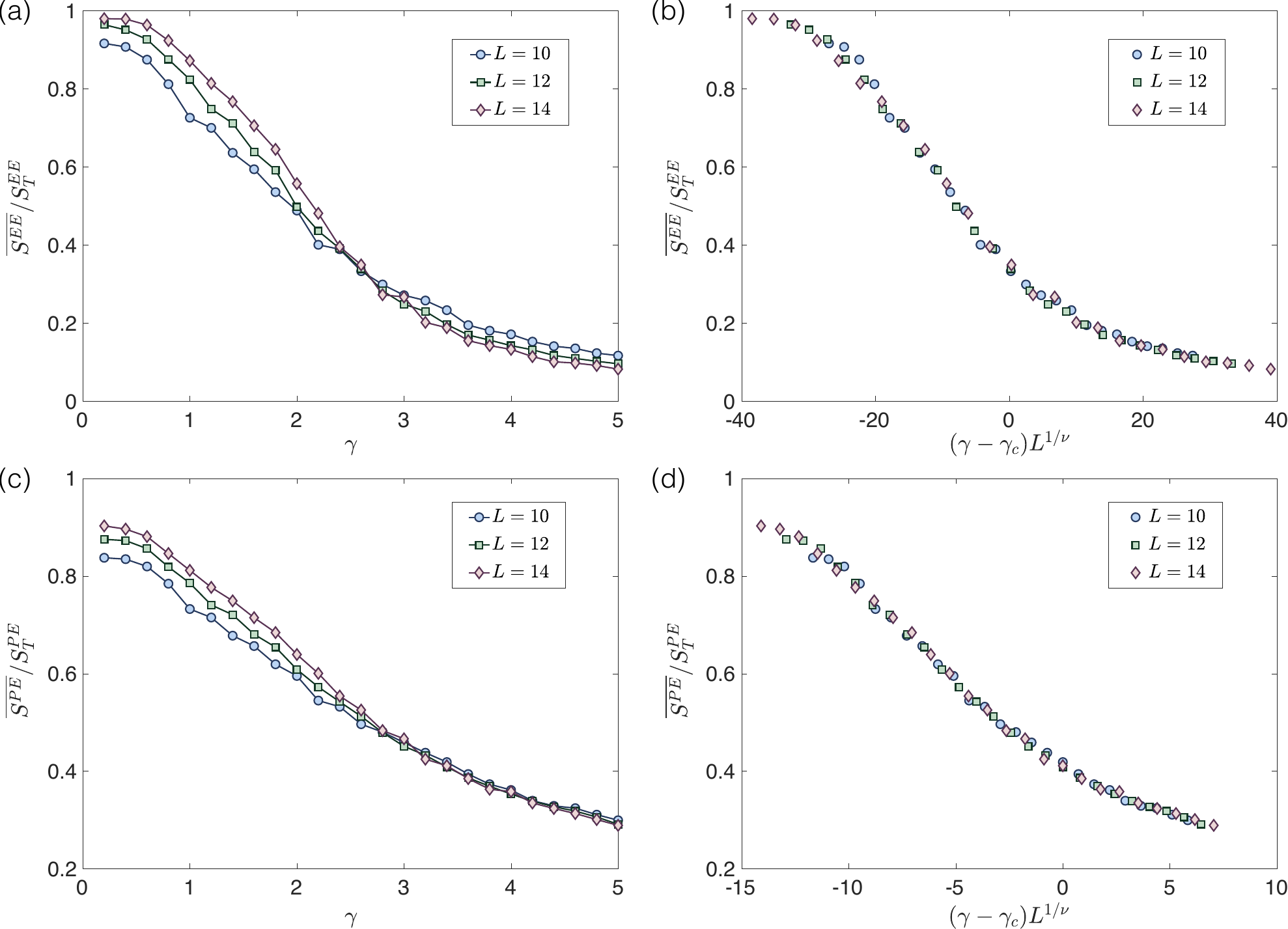}\\
	\caption{(a) The half-chain EE, divided by the Page value $S^{EE}_{T}$, as a function of $\gamma$ in the Hamiltonian (\ref{H_bh}) with $\alpha=2$, $U=4$, $J=1$, and system sizes $L=10$, $12$ and $14$. (b) Finite-size critical scaling collapse for the EE data. The critical strength of linear potential $\gamma_{c}\simeq 2.58$ and the scaling exponent $\nu\simeq 0.949$ are used to the most appropriate data collapse. (c) is similar to (a) but for the PE divided by $S^{PE}_{T} = \ln \mathcal{N}$. (d) Finite-size critical scaling collapse for the PE data. The critical strength of linear potential $\gamma_{c}\simeq 3.40$ and the scaling exponent $\nu\simeq 1.78$ are used to the most appropriate data collapse. }\label{fig3}
\end{figure*}

The Bose-Hubbard model, describing a 1D array of superconducting
qubits, reads~\cite{AQS1,AQS2,AQS5,BH_add2,BH_add3}
\begin{eqnarray} \nonumber
\hat{H}_{1} = &-& \frac{U}{2}\sum_{j=0}^{L-1} \hat{n}_{j}(\hat{n}_{j} - 1) + \sum_{j=0}^{L-1}h_{j} \hat{n}_{j} \\
 &+& J \sum_{j=0}^{L-2} (\hat{a}_{j}^{\dagger}\hat{a}_{j+1} + \text{H.c.}),
\label{H_bh}
\end{eqnarray}
where $\hat{n}_{j}=\hat{a}_{j}^{\dagger}\hat{a}_{j}$ is the bosonic number operator with $\hat{a}_{j}^{\dagger}$ ($\hat{a}_{j}$) being the corresponding creation (annihilation) operator, $J$ refers to the hopping interaction between the $j$-th and $(j+1)$-th qubit, $U$ denotes the qubit anharmonicity, serving as the on-site interaction, and $h_{j}$ is the chemical potential of $j$-th qubit. Due to the conservation of total number of bosonic excitations, we can study the Hamiltonian (\ref{H_bh}) with the half-filling subspace.

To study the Stark many-body localization transition, we consider the on-site linear potential~\cite{stark1,stark5}
\begin{eqnarray}
h_{j} = -\gamma j +\alpha (\frac{j}{L-1})^{2}
\label{linear}
\end{eqnarray}
with $\gamma$ as the strength of the potential, and $\alpha$ being a parameter that breaks the pure linearity. The non-zero $\alpha$ is required for a direct comparison to the MBL induced by disorder fields, i.e., the conventional MBL. A scheme for the Bose-Hubbard model (\ref{H_bh}), in the presence of the linear potential with curvature, i.e., Eq.~(\ref{linear}), is plotted in Fig.~\ref{fig1}.

\subsection{Results}

\begin{figure}
	\centering
	\includegraphics[width=1\linewidth]{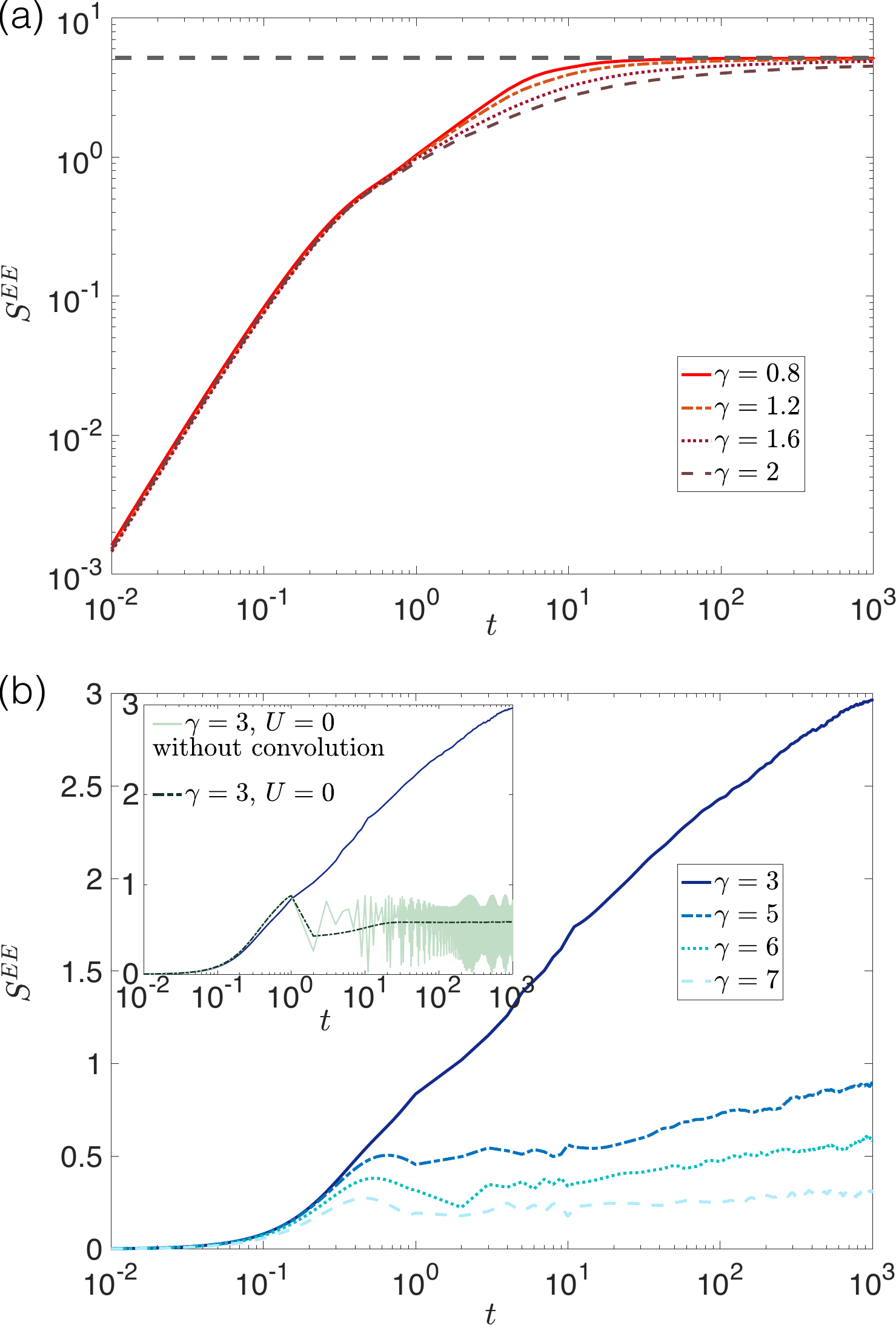}\\
	\caption{(a) The dynamics of EE in the Hamiltonian (\ref{H_bh}) with $L=14$, $J=1$, $U=4$, $\alpha=2$, and several strengths of linear potential $\gamma$ smaller than the critical value. (b) The dynamics of EE in the same system in (a) but for larger $\gamma$. The inset of (b) presents the dynamics of EE with $\gamma=3$ and $U=4$, in comparison with the noninteracting condition $U=0$. All data have been smoothed by the convolution (similar to Ref.~\cite{stark1}), except for the light green line shown in the inset of (b). The horizontal dashed line in (a) marks the Page value.}\label{fig4}
\end{figure}

\begin{figure}
	\centering
	\includegraphics[width=1\linewidth]{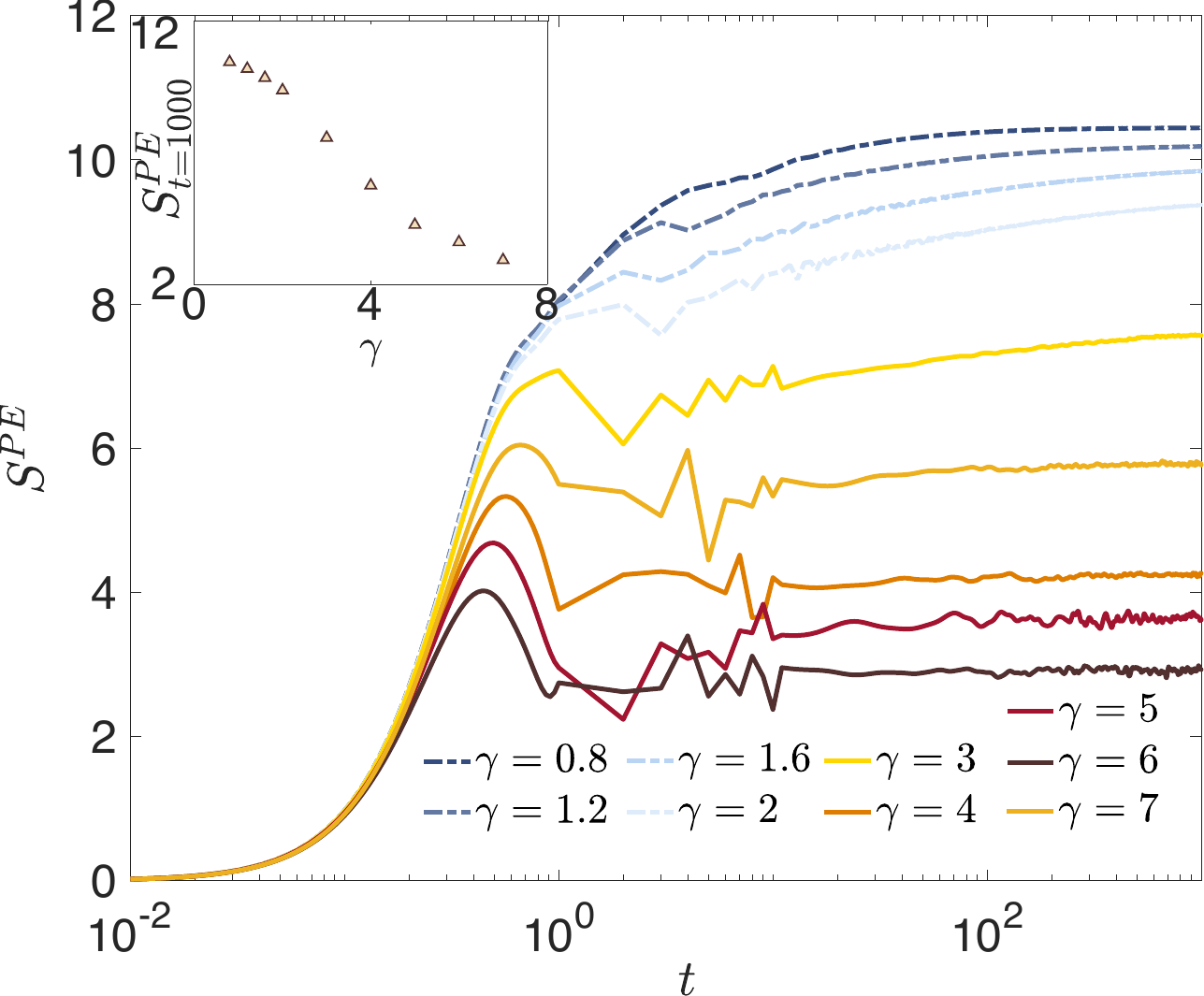}\\
	\caption{The dynamics of PE in the Hamiltonian (\ref{H_bh}) with $L=14$, $J=1$, $U=4$, $\alpha=2$, and several strengths of linear potential $\gamma$. The inset shows the value of PE at $t=1000$ as a function of $\gamma$. All data have been smoothed by the convolution. }\label{fig5}
\end{figure}

An efficient way to characterize the ergodicity and MBL relies on studying the level statistics based on random matrix theory~\cite{rog1}. The ratio of consecutive level spacings is defined as
\begin{eqnarray}
r^{(n)} = \frac{\min[\delta^{(n)}, \delta^{(n+1)}]}{\max[\delta^{(n)}, \delta^{(n+1)}]}
\label{rog}
\end{eqnarray}
with $\delta^{(n)} = E_{n}-E_{n-1}>0$ being the difference between the $n$-th and $(n-1)$-th eigenenergy of the Hamiltonian (\ref{H_bh}). The $\langle r\rangle $, being the $r^{(n)}$ averaged over the eigenenergies in a chosen window, has been employed to study the conventional MBL~\cite{AQS5,MBL_num1,MBL_num2,MBL_num5,BH_MBL1,BH_MBL2,BH_MBL3,BH_MBL4,rog2,rog3,rog4}, and the Stark MBL~\cite{stark1,stark2,stark5}, showing that the $r^{(n)}$ of a system in the ergodic and many-body localized phase satisfy the Wigner-Dyson ($\langle r\rangle \simeq 0.5307$) and Poisson distribution ($\langle r\rangle \simeq 0.3863$), respectively.

Before we calculate the $r^{(n)}$, considering the influence of the many-body mobility edge (MBME)~\cite{AQS4,MBL_num2,MBL_num3,stark_a1,mbme_a1,mbme_a2,mbme_a3,mbme_a4}, we should first chose an energy window around the energy with the largest density of state (DoS) (see Appendix A for details). For a highly-excited eigenstate, the normalized energy is defined as
\begin{eqnarray}
 \epsilon = \frac{E-E_{\text{min}}}{E_{\text{max}}-E_{\text{min}}}
\label{normalized_energy}
\end{eqnarray}
with $E_{\text{max}}$ ($E_{\text{min}}$) as the largest (smallest) eigenenergys of the Hamiltonian, and $E$ being the corresponding eigenenergy of the highly-excited eigenstate. The energy window, adopted for the calculation of $\langle r\rangle$, is $\epsilon \in [\epsilon^{*}-0.05,\epsilon^{*}+0.05]$, where $\epsilon^{*}$ is the normalized energy with the largest DoS.

Using the ED method, we calculate the $\langle r\rangle$ in the Hamiltonian (\ref{H_bh}) with the system size $L=12$, $\alpha=2$, and various $\gamma$ and $U$. The results are presented in Fig.~\ref{fig2}. It has been shown that $\langle r\rangle\simeq 0.44$ around the critical point of the Stark MBL transition~\cite{stark2}. Thus, we mark the contour for $\langle r\rangle\simeq 0.44$ in Fig.~\ref{fig2}. One can see that the maximum critical strength of the linear potential is reached when $U\simeq 4$, which is consistent with the conventional MBL in the 1D Bose-Hubbard model~\cite{BH_MBL1}. Below, we mainly focus on the Stark MBL transition in the Bose-Hubbard model (\ref{H_bh}) with $J=1$ and $U=4$.

To more accurately estimate the critical strength of the linear potential $\gamma_{c}$, we perform the finite-size scaling analysis of the EE and PE for the highly-excited eigenstates $|E\rangle$ with the normalized energy $\epsilon \in [\epsilon^{*}-0.05,\epsilon^{*}+0.05]$. The definition of EE is
\begin{eqnarray}
S^{EE} = -\text{Tr} (\hat{\rho}_{\mathcal{A}} \ln \hat{\rho}_{\mathcal{A}} ),
\label{EE}
\end{eqnarray}
where $\hat{\rho}_{\mathcal{A}} = \text{Tr}_{\mathcal{B}}(|E\rangle\langle E|)$ is a reduced density matrix of the subsystem $\mathcal{A}$, and the remainder is the subsystem $\mathcal{B}$. Here, we consider a half-chain subsystem, which consists of the qubit $Q_{0}$, $Q_{1}$, ..., and $Q_{\frac{L}{2}-1}$. Moreover, the PE reads
\begin{eqnarray}
S^{PE} = -\sum_{i}^{\mathcal{N}} p_{i}\ln p_{i},
\label{PE}
\end{eqnarray}
where $\mathcal{N}$ is the dimension of Hilbert space, and $p_{i} = |\langle E| i\rangle|^{2}$ with a basis $|i\rangle$. Here, we chose the basis
\begin{eqnarray}
|i\rangle \in \{|n_{1}n_{2}...n_{L}\rangle| \sum_{j=1}^{L} n_{j} = \frac{L}{2}, n_{j}\in \mathbb{N}, 0\leq n_{j} \leq \frac{L}{2}\}.
\label{basis}
\end{eqnarray}

\begin{figure*}
	\centering
	\includegraphics[width=1\linewidth]{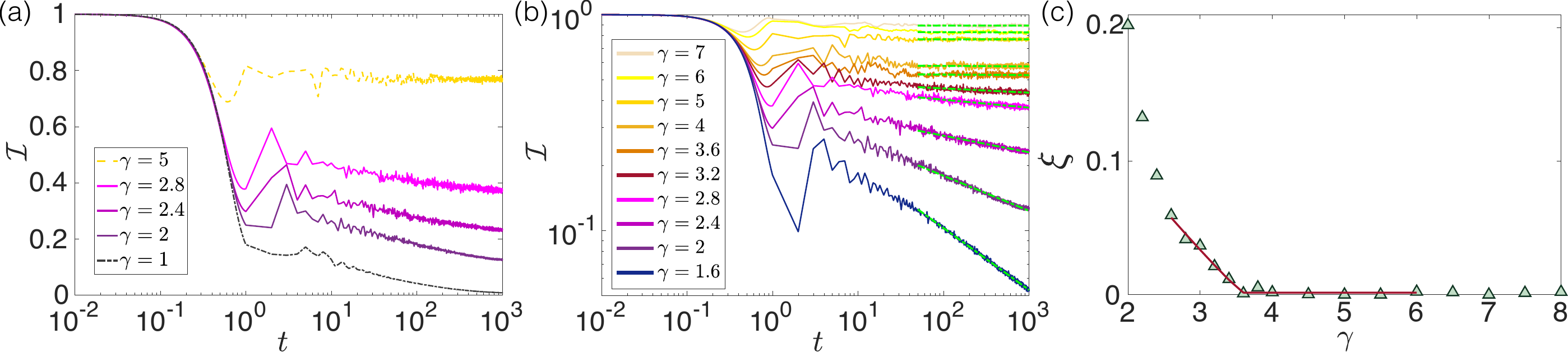}\\
	\caption{(a) The dynamics of imbalance in the Hamiltonian (\ref{H_bh}) with $L=14$, $J=1$, $U=4$, $\alpha=2$, and several strengths of linear potential $\gamma$ on a logarithmic $x$ axis. (b) is similar to (a) but with different $\gamma$ and on a double-log plot. (c) The exponent of imbalance decay $\xi$ as a function of $\gamma$. The dashed lines in (b) are the power-law fit $\mathcal{I}(t)\propto t^{-\xi}$. The solid line in (c) is a piecewise fit $\xi\propto\{|\gamma-\gamma_{c}|^{\nu} + \overline{\xi},\overline{\xi}\}$ in the regime $2\leq \gamma \leq 8$, yielding $\gamma_{c}\simeq 3.59$ and $\nu\simeq 1.08$.   }\label{fig6}
\end{figure*}

The results of EE and PE averaged over the eigenstates $|E\rangle$, i.e., $\overline{S^{EE}}$ and $\overline{S^{PE}}$, are displayed in Fig.~\ref{fig3}(a) and (c), respectively. The $\overline{S^{EE}}$ is divided by the Page value~\cite{Page} for a random pure state $S^{EE}_{T}$ (see Appendix B). Moreover, the $\overline{S^{PE}}$ is normalized by $S^{PE}_{T} = \ln \mathcal{N}$~\cite{MBL_num6}. For both the normalized EE and PE, the crossings can be observed for the data with different $L$. The crossings of $\overline{S^{EE}}/S^{EE}_{T}$, around $\gamma\simeq 2.5$, are more obvious than those of $\overline{S^{PE}}/S^{PE}_{T}$. We then perform the scaling data collapse of $\overline{S^{EE}}/S^{EE}_{T}$ and $\overline{S^{PE}}/S^{PE}_{T}$ by employing the ansatz $f[L^{1/\nu}(\gamma-\gamma_{c})]$. The results after the data collapse are presented in Fig.~\ref{fig3}(b) and (d).

Two points of note. First, the critical point estimated by the data of EE is $\gamma_{c}^{EE} \simeq 2.58$. However, the data collapse of PE gives $\gamma_{c}^{PE} \simeq 3.40$. For the conventional MBL, it has been suggested that the critical point obtained from the scaling analysis of EE is regarded as a \emph{lower bound} of the critical point~\cite{MBL_num1,MBL_num_a2}, supporting $\gamma_{c}^{PE} \geq \gamma_{c}^{EE}$. Second, the scaling exponent extracted from the data collapse of EE is $\nu^{EE}\simeq 0.949$. Similar to previous numerical works studying the EE using the ED method, the scaling exponent violates the Harris-Chayes-Chayes-Fisher-Spencer (Harris-CCFS) bound $\nu>2/d$~\cite{bound} ($d=1$ being the dimension of the system). Nevertheless, we see that the data collapse of PE yields $\nu^{PE}\simeq 1.78$, which is closer to the Harris-CCFS bound than $\nu^{EE}$.

Moreover, we also employ the infinite-order Berezinskii-Kosterlitz-Thouless (BKT) type of transition~\cite{BKT1} to perform the finite-size critical scaling collapse for the data shown in Fig.~\ref{fig3}(a) and (c). Although it has been suggested that the scaling properties of the conventional MBL transition better agree with the BKT scenario than the ansatz used for the data collapse in Fig.~\ref{fig3}(b) and (d)~\cite{BKT2,BKT3}, we numerically show that the BKT scenario does not significantly improve the quantity of the data collapse for the Stark MBL transition in the Bose-Hubbard model (\ref{H_bh}) (see Appendix C for details).

Next, we numerically study the non-equilibrium properties of the Hamiltonian (\ref{H_bh}) with the linear potential (\ref{linear}) by using the Krylov subspace methods~\cite{stark2,BH_MBL1}. The choice of initial states $|\psi_{0}\rangle$ can influence the signatures of MBL transition because of the MBME~\cite{AQS4,mbme_a4}. The normalized energy of an initial state can be characterized by
\begin{eqnarray}
\epsilon(|\psi_{0}\rangle) = \frac{\langle\psi_{0}|\hat{H}|\psi_{0}\rangle-E_{\text{min}}}{E_{\text{max}}-E_{\text{min}}},
\label{normalized_energy_0}
\end{eqnarray}
where $E_{\text{max}}$ ($E_{\text{min}}$) is the largest (smallest) eigenenergys of the Hamiltonian $\hat{H}$. Here, we chose the initial state $|\psi_{0}\rangle$ whose normalized energy satisfies  $\epsilon(|\psi_{0}\rangle) \in [\epsilon^{*}-0.05,\epsilon^{*}+0.05]$. In addition, previous quantum simulations of MBL mainly consider a product state $|n_{1}n_{2}...n_{L}\rangle$ with $n_{j}\in\{0,1\}$~\cite{AQS4,cold_atom1,cold_atom2,cold_atom3,cold_atom4,cold_atom5,tapped_ion1,tapped_ion2}. Consequently, besides the constraint about the normalized energy, we consider the initial state
\begin{eqnarray}
|\psi_{0}\rangle \in \{|n_{1}n_{2}...n_{L}\rangle| \sum_{j=1}^{L} n_{j} = \frac{L}{2}, n_{j}\in \mathbb{N}, 0\leq n_{j} \leq 1\}.
\label{psi_0}
\end{eqnarray}

We first calculate the time evolution of the EE $S^{EE}$ for the quenched state $|\psi_{t}\rangle = \exp(-i\hat{H}_{1} t)|\psi_{0}\rangle$ with different strength of the linear potential $\gamma$. The results of $S^{EE}$ averaged over 50 initial states are shown in Fig.~\ref{fig4}. As shown in Fig.~\ref{fig4}(a), with $\gamma<\gamma_{c}$, the EE exhibits a ballistic spreading $S^{EE}(t)\propto t^{\beta}$ under its initial evolution~\cite{EE_thermal}, and then reaches the Page value~\cite{Page}, identifying the ergodic dynamics. With larger $\gamma$, the logarithmic growth of EE is observed in Fig.~\ref{fig4}(b), which is absent for the noninteracting Stark localization in free bosonic system, i.e., the Hamiltonian (\ref{H_bh}) with $U=0$ [see the inset of Fig.~\ref{fig4}(b)]. In addition, the results of the EE dynamics with larger $U$ are presented in Appendix D, showing that with the increase of $U$, the Stark MBL tends to its noninteracting counterpart.

Although the EE is a powerful tool to characterize the Stark MBL out-of-equilibrium, the measurement of EE, using the quantum state tomography, remains an experimental challenge, especially for a multi-qubit subsystem. As discussed in Ref.~\cite{MBL_num_dy4}, the measurement of PE defined in Eq.~(\ref{PE}) is more efficient than that of the half-chain EE. Thus, we also study the quench dynamics of PE, and plot the results in Fig.~\ref{fig5}. It is shown that with the increase of $\gamma$, the growth of PE becomes slower. Moreover, in Appendix C, we observe that when $\gamma=2$, the growth of PE shown in Fig.~\ref{fig5} is suppressed for the noninteracting case.

Besides the dynamics of EE and PE, the imbalance can also be employed to study the conventional MBL, both numerically~\cite{MBL_num_a2,MBL_num_dy1,MBL_num_dy2,MBL_num_dy3} and experimentally~\cite{AQS2,AQS4,cold_atom1,cold_atom2,cold_atom3}. For the initial state in Eq.~(\ref{psi_0}), the imbalance is defined as
\begin{eqnarray}
\mathcal{I} = \frac{N_{1}-N_{0}}{N_{1}+N_{0}},
\label{imbalance}
\end{eqnarray}
where $N_{1(0)} =2 \sum_{j,|n_{j}\rangle=|1 (0)\rangle} \langle\psi_{t} |\hat{n}_{j}|\psi_{t} \rangle/L$. In the ergodic phase, the imbalance $\mathcal{I}=0$ when $t\rightarrow \infty$. For the Stark MBL, it has been shown that the value of $\mathcal{I}$ becomes finite and tends to $1$ with the increase of $\gamma$~\cite{stark1,stark2,stark5}. Previous studies of the disorder-induced MBL have revealed a regime where the imbalance exhibits a slow relaxation at the ergodic side of the MBL transition~\cite{cold_atom3, Griffiths1,Griffiths2,Griffiths3,Griffiths4}.
However, the relaxation of imbalance for the Stark MBL has not been carefully investigated so far.

The results of imbalance for the Hamiltonian (\ref{H_bh}) in the presence of a linear potential (\ref{linear}) are displayed in Fig.~\ref{fig6}. As shown in Fig.~\ref{fig6}(a), three dynamical regimes are observed. The first regime is at a weak strength of linear potential (for instance, $\gamma=1$), where the imbalance exhibits a quick relaxation, approaching to 0 at a long time. The second one is a regime of slow relaxation with intermediate strength of linear potential [see the results of $\gamma=2$, $2.4$ and $2.8$ in Fig.~\ref{fig6}(a)]. The third one is the regime of Stark many-body localized phase with negligible relaxation (for instance, $\gamma=5$). In Fig.~\ref{fig6}(b), we show that the slow relaxation of imbalance is dominated by a power-law decay $\mathcal{I}\propto t^{-\xi}$.

As a side remark, in Appendix E, we explore the impact of the parameter $\alpha$ in Eq.~(\ref{linear}) on the time evolution of imbalance. With large strength of the linear potential $\gamma>\gamma_{c}$, when $\alpha=0$, the dynamical behavior of imbalance is more sensitive to the choice of initial states than that with finite $\alpha$, and the quench dynamics may fail to retain the memory of initial states. One can see more details and discussions in Appendix E.


The exponent $\xi$, extracted from the relaxation of $\mathcal{I}(t)$, can provide valuable insights into the phase transition between the ergodic and many-body localized phase. It has been theoretically suggested that the exponent $\xi$ satisfies
\begin{equation}
\xi\propto
 \begin{cases}
 |W-W_{c}|^{\nu} & (\gamma<\gamma_{c})\\
 0  & (\gamma\geq\gamma_{c})
 \end{cases},
 \label{imbalance_critical}
\end{equation}
where $W_{c}$ and $\nu$ denote the critical strength disorder and the scaling exponent for the conventional MBL, respectively~\cite{imbalance_critical}, which has also been experimentally verified~\cite{cold_atom3}. Nevertheless, for the Stark MBL transition, the relation (\ref{imbalance_critical}) has not yet been demonstrated. In Fig.~\ref{fig6}(c), we present the exponent $\xi$ as a function of $\gamma$, and fit the data using $\xi\propto\{|\gamma-\gamma_{c}|^{\nu}+\overline{\xi},\overline{\xi}\}$. The $\overline{\xi}$ denotes the exponent $\xi$ averaged over the data with $\gamma\geq 3.6$, which is a finite value near $0$, taking the numerical error into consideration. We then obtain the critical strength of linear potential $\gamma_{c}^{I}\simeq 3.8$ and $\nu\simeq 1.85$. The $\gamma_{c}^{I}$ extracted from the data of imbalance is much larger than the $\gamma_{c}$ obtained from the calculation of the EE and PE of highly-excited eigenstates [see Fig.~\ref{fig2}(b) and (d)].


\begin{figure}
	\centering
	\includegraphics[width=1\linewidth]{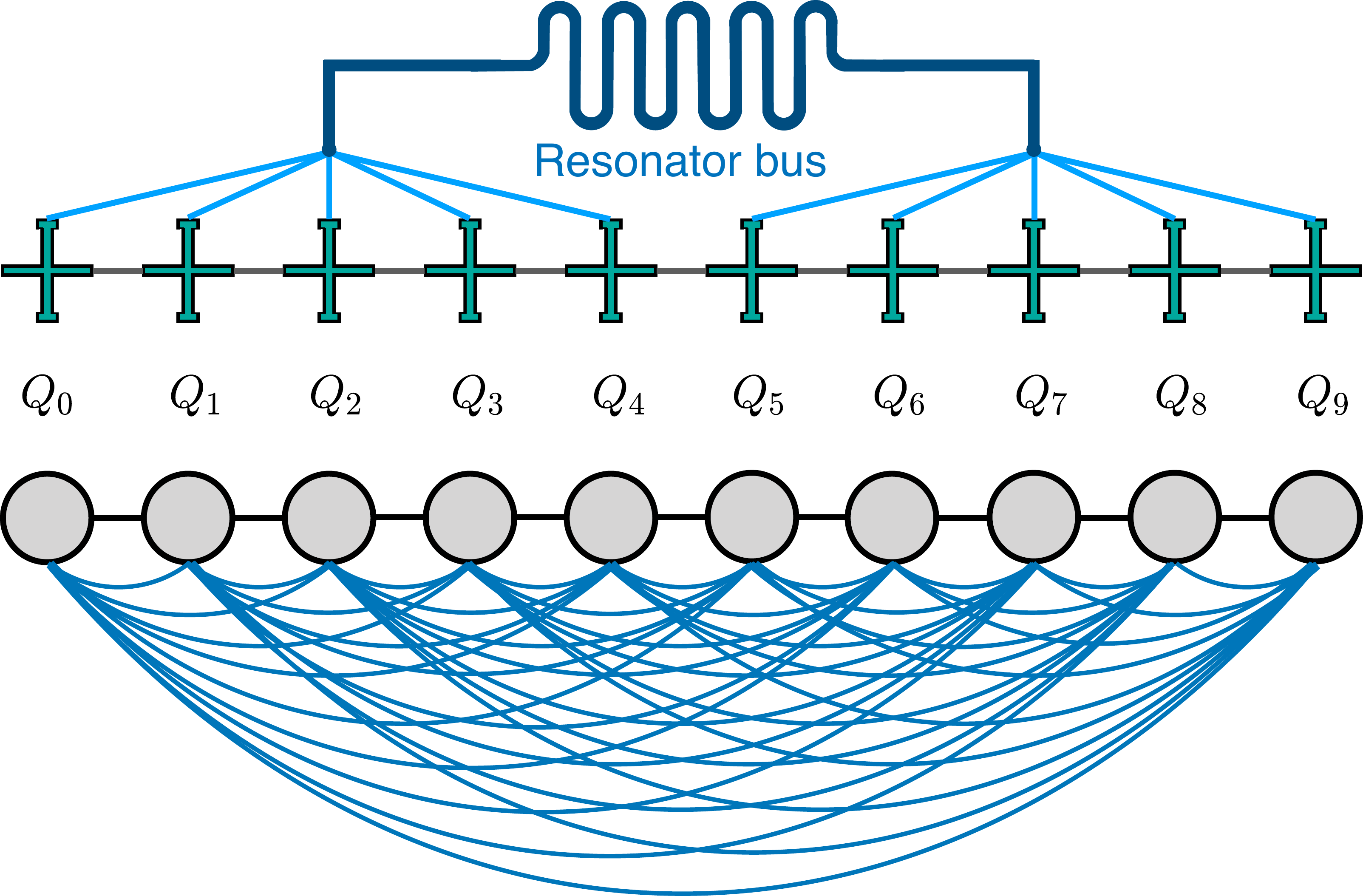}\\
	\caption{Top panel: A schematic illustration of the superconducting circuit with all-to-all connectivity. Besides the nearest-neighbor couplings in the 1D array, the non-local interactions are enabled by the resonator bus, coupling to each individual qubits. Bottom panel: A sketch of the hopping interactions between qubits. The black lines and blue lines present the nearest-neighbor interactions and non-local interactions mediated by the resonator bus, respectively.}\label{fig7}
\end{figure}

\section{The Stark many-body localization transition in the superconducting circuit with all-to-all connectivity}
\subsection{Model}

\begin{figure}
	\centering
	\includegraphics[width=1\linewidth]{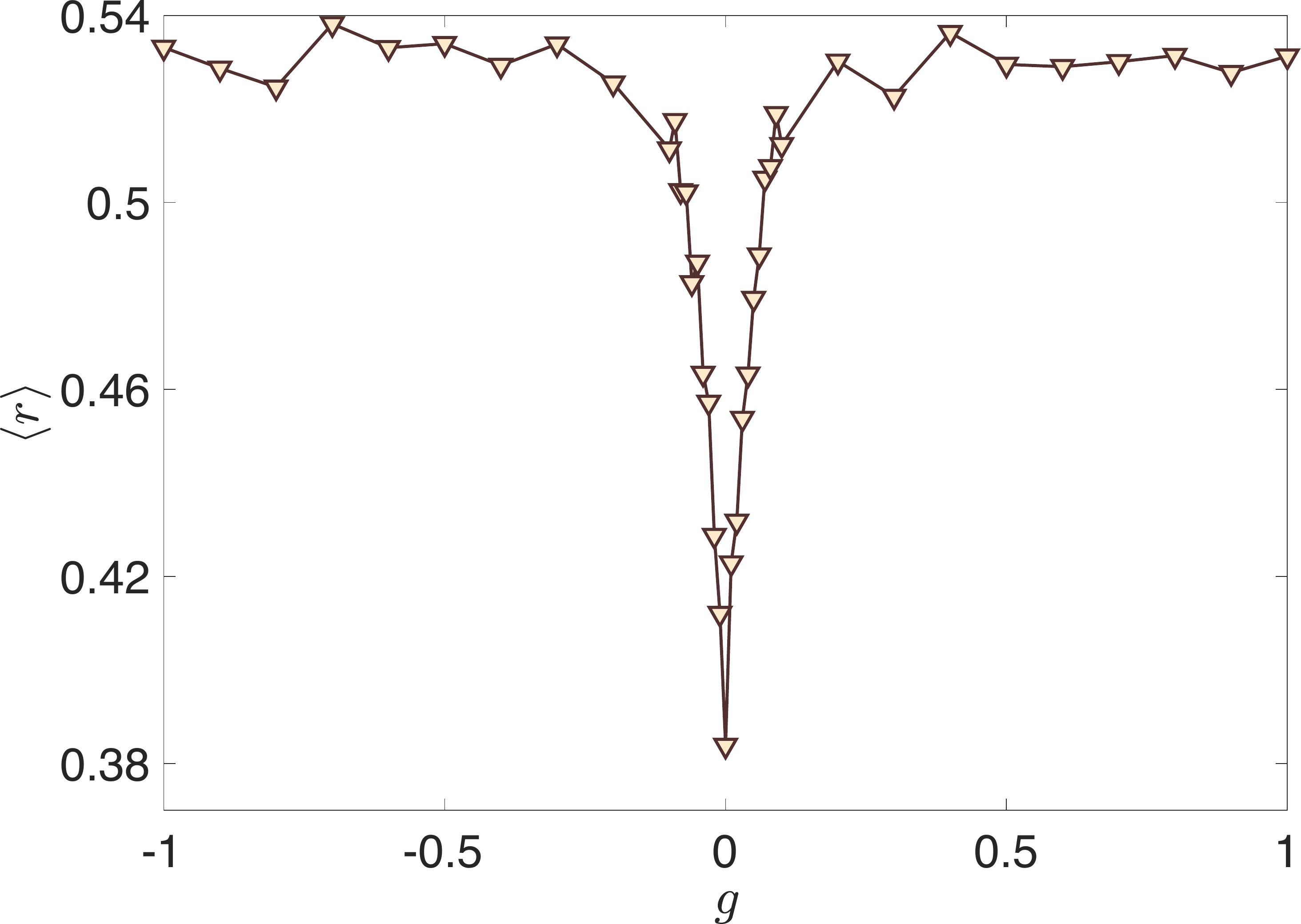}\\
	\caption{The ratio of consecutive level spacings $\langle r\rangle $ averaged over a chosen energy window in the Hamiltonian~(\ref{H_all_to_all}) with $L=16$, $\alpha=2$, and $\gamma=0.2$, as a function of $g$.}\label{fig8}
\end{figure}

\begin{figure*}
	\centering
	\includegraphics[width=0.9\linewidth]{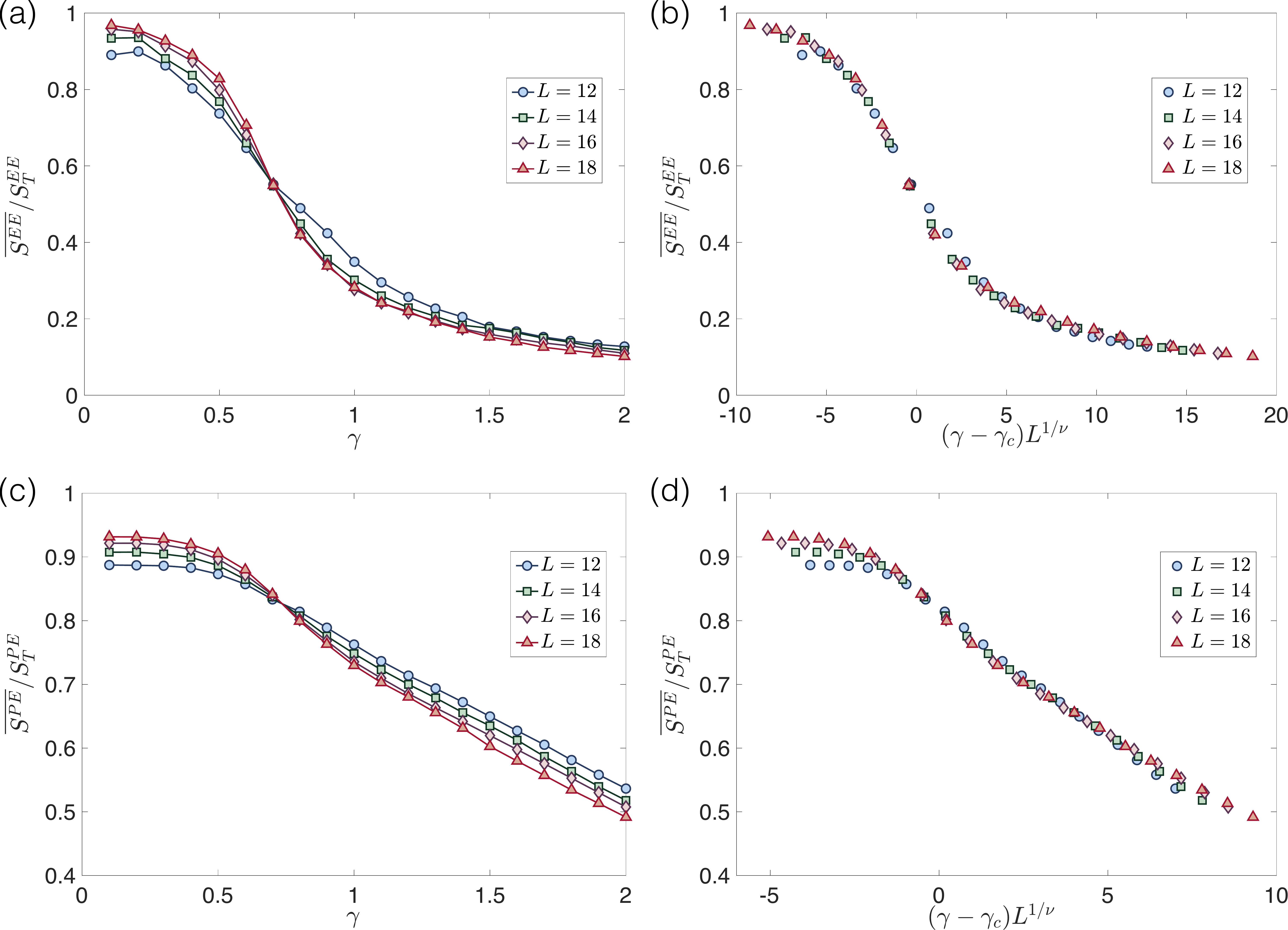}\\
	\caption{(a) The half-chain EE, divided by the Page value $S_{T}^{EE}$, as a function of $\gamma$ in the Hamiltonian (\ref{H_all_to_all}) with $\alpha=2$, $g=0.5$, and system sizes $L=12$, $14$, $16$, and $18$. (b) Finite-size critical scaling collapse for the EE data. The critical strength of linear potential $\gamma_{c}\simeq 0.73$ and the scaling exponent $\nu\simeq 1.07$ are used to the most appropriate data collapse. (c) is similar to (a) but for the PE divided by $S^{PE}_{T} = \ln \mathcal{N}$. (d) Finite-size critical scaling collapse for the PE data. The critical strength of linear potential $\gamma_{c}\simeq 0.77$ and the scaling exponent $\nu\simeq 1.43$ are used to the most appropriate data collapse.}\label{fig9}
\end{figure*}

The sign of MBL has been observed in the superconducting circuit with all-to-all connectivity~\cite{AQS4,AQS_a1}. As shown in Fig.~\ref{fig7}, the interactions between qubits consist of two parts. The first part is the nearest-neighbor crosstalk couplings between the qubit $Q_{i}$ and $Q_{i+1}$. The second one is the resonator-induced long-range couplings. For the superconducting circuit with all-to-all connectivity, for instance the devices in Refs.~\cite{AQS4,AQS_a1} , the ratio of qubit anharmonicity $U$ to the hopping interaction $J$ is $U/J\sim 10^{2}$. Hence, the nearest-neighbor bosonic hopping interaction $\sum_{j}^{L-1}(\hat{a}^{\dagger}_{j}\hat{a}_{j+1} + \text{H.c.})$ approximates to the hard-core limit~\cite{BH_add}, i.e., $\sum_{j}^{L-1}(\hat{\sigma}^{+}_{j}\hat{\sigma}^{-}_{j+1} + \text{H.c.}) \propto \hat{H}_{XX} \equiv \sum_{j}^{L-1}(\hat{\sigma}^{x}_{j}\hat{\sigma}^{x}_{j+1} + \hat{\sigma}^{y}_{j}\hat{\sigma}^{y}_{j+1})$, being the $XX$ interaction, with $\hat{\sigma}^{\alpha}$ ($\alpha\in\{x,y,z\}$) as the Pauli matrix (see more results in Appendix D). The resonator bus enables the long-range couplings between two qubits, which do not decay with the increase of distance, i.e., $\hat{H}_{\text{non-local}} = \sum_{i<j}(\hat{\sigma}^{x}_{i}\hat{\sigma}^{x}_{j} + \hat{\sigma}^{y}_{i}\hat{\sigma}^{y}_{j})$. In short, the Hamiltonian of the all-to-all connected superconducting qubits reads
\begin{eqnarray} \nonumber
\hat{H}_{2} = &-&\frac{g}{L}\sum_{j>i+1}^{L-1} (\hat{\sigma}^{x}_{i}\hat{\sigma}^{x}_{j} + \hat{\sigma}^{y}_{i}\hat{\sigma}^{y}_{j}) \\
&+& \sum_{i=0}^{L-2} (\hat{\sigma}^{x}_{i}\hat{\sigma}^{x}_{i+1} + \hat{\sigma}^{y}_{i}\hat{\sigma}^{y}_{i+1}) +\sum_{i=0}^{L-1} h_{i}\hat{\sigma}^{z}_{i},
\label{H_all_to_all}
\end{eqnarray}
where $L$ is the number of qubits, and $h_{i}$ is the on-site potential in Eq.~(\ref{linear}). Note that the Hamiltonian (\ref{H_all_to_all}) has a conservation of $\sum_{i=0}^{L-1}\hat{\sigma}^{z}_{i}$. As a direct consequence, we focus on the highly-excited eigenstates in the half-filling sector, and for the quench dynamics, we chose the initial state $|\psi_{0}\rangle$ satisfying $\langle\psi_{0}|\sum_{i=0}^{L-1}\hat{\sigma}^{z}_{i}|\psi_{0}\rangle=0$.

In the system (\ref{H_all_to_all}), when $g = h_{i} = 0$, $\hat{H}_{2}$ is an integrable model~\cite{free_XY}. To characterize the transition from integrability to the ergodic phase, we study the ratio of consecutive level spacings $\langle r\rangle $ in the window of normalized energy $\epsilon\in [0.48,0.52]$ (as shown in Appendix A, the normalized energy with the maximum DoS is $\epsilon^{*}\simeq 0.5$). The results are presented in Fig.~\ref{fig8}. In the vicinity of $g \simeq 0$, it is seen that $\langle r\rangle \simeq 0.3863$, agreeing with the Poisson distribution. In contrast, $\langle r\rangle \simeq 0.5307$, for the Wigner-Dyson distribution, is observed when $|g|\geq 0.3$. Consequently, to study the Stark MBL transition in the Hamiltonian (\ref{H_all_to_all}), we consider $g=0.5$.

\subsection{Results}

\begin{figure}
	\centering
	\includegraphics[width=1\linewidth]{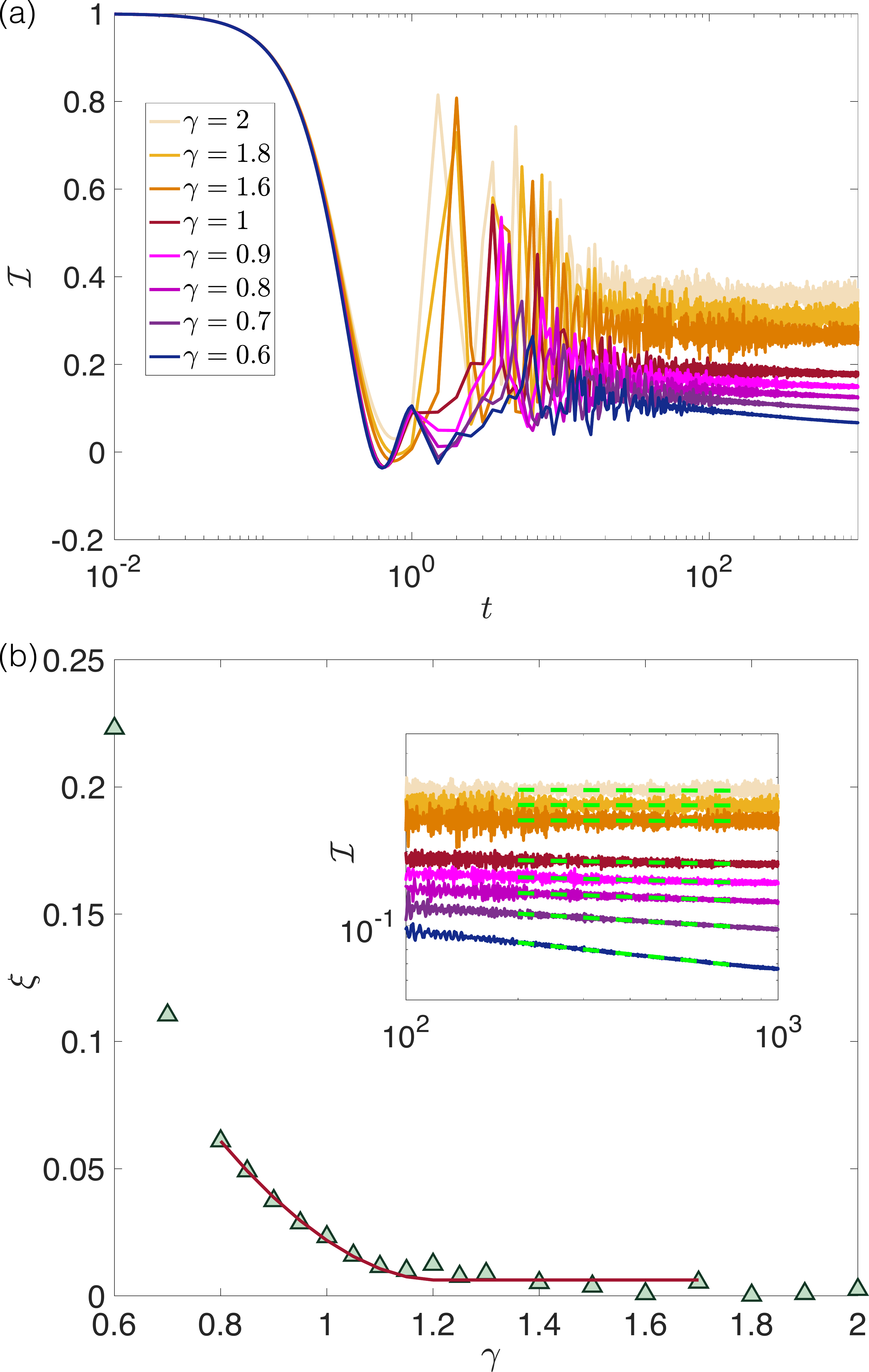}\\
	\caption{(a) Time evolution of imbalance in the Hamiltonian (\ref{H_all_to_all}) with $L=18$, $g=0.5$, $\alpha=2$, and several strengths of linear potential $\gamma$ on a logarithmic $x$ axis. (b) The exponent of imbalance decay $\xi$ as a function of $\gamma$. The solid line in (b) is a piecewise fit $\xi\propto\{|\gamma-\gamma_{c}|^{\nu}+\overline{\xi},\overline{\xi}\}$ in the regime $0.8\leq \gamma \leq 2$, yielding $\gamma_{c}\simeq 1.20$ and $\nu\simeq 1.81$. The inset of (b) shows the dynamics of imbalance on a double-log plot. The dashed lines in the inset of (b) are the power-law fit $\mathcal{I}(t)\propto t^{-\xi}$. }\label{fig10}
\end{figure}

We now study the EE and PE for the highly-excited eigenstates $|E\rangle$ with the normalized energy $\epsilon\in[0.48,0.52]$, and employ similar finite-size critical scaling collapse in Fig.~\ref{fig3} for the data of EE and PE with different system size $L$ to estimate the critical value of linear potential $\gamma_{c}$. The results are plotted in Fig.~\ref{fig9}. The scaling analysis of the EE gives a lower bound of the critical linear potential $\gamma_{c}^{EE}\geq 0.73$. The $\gamma_{c}$ obtained from the data of PE is $\gamma_{c}^{PE}\simeq 0.77 > \gamma_{c}^{EE}$.

Next, we study the time evolution of imbalance with the initial states $|\psi_{0}\rangle$ defined in (\ref{psi_0}), whose normalized energies satisfy $\epsilon(|\psi_{0}\rangle)\in[0.48,0.52]$. As shown in Fig.~\ref{fig10}(a), the slow relaxation of the imbalance $\mathcal{I}$ can also be observed in the system (\ref{H_all_to_all}). We fit the long-time behavior of $\mathcal{I}$ ($t\geq 200$) by power-law decay $t^{-\xi}$ [see the inset of Fig.~\ref{fig10}(b)], and display the decay exponent $\xi$ as a function of $\gamma$ in Fig.~\ref{fig10}(b). Taking the ansatz $\xi\propto\{|\gamma-\gamma_{c}|^{\nu} + \overline{\xi},\overline{\xi}\}$ (here the $\overline{\xi}$ is the exponent averaged over the data with $\gamma\geq 1.2$) into consideration, we obtain the scaling exponent $\nu\simeq 1.1$, and $\gamma_{c}^{I}\simeq 1.1$, which is larger than the $\gamma_{c}$ in Fig.~\ref{fig9}(b) and (d).

Employing the scaling analysis of the EE and PE of highly-excited eigenstates, and the exponent $\xi$ extracted from the slow relaxation of imbalance, we obtain three estimated values of the critical linear potential $\gamma_{c}^{EE}$, $\gamma_{c}^{PE}$, and $\gamma_{c}^{I}$. For both the Hamiltonian (\ref{H_all_to_all}) and (\ref{H_bh}), the results suggest that $\gamma_{c}^{I}>\gamma_{c}^{PE}>\gamma_{c}^{EE}$.

Moreover, we study the dynamics of EE and PE in the system (\ref{H_all_to_all}). The results and related discussions are displayed in Appendix F. In the Stark many-body localized phase, the behavior of PE is similar to that in the system (\ref{H_bh}) (see Fig.~\ref{fig5} and Fig.~\ref{fig:S6}), while the dynamics of EE satisfies a power-law growth. The absence of logarithmic growth of EE in many-body localized phase is also observed in disordered spin chains with power-law interactions~\cite{Safavi-Naini2019}.



\section{Discussion}

In this paper, we have systematically explored the Stark MBL transition in a chain of superconducting qubits, described by the 1D Bose-Hubbard model, and the superconducting circuit with all-to-all connectivity, comprised of qubits with large anharmonicity. Based on the finite-size scaling collapse of the EE and PE for highly-excited eigenstates, we obtain the lower bound of the critical linear potential $\gamma_{c}$ for the Stark MBL transition. By studying the out-of-equilibrium dynamics in the Stark many-body localized phase, we observe slow growth of PE, logarithmic growth of EE, and the nonzero stationary value of imbalance.

The slow dynamical behavior of imbalance, well fitted by a power-law decay $\mathcal{I}(t) \propto t^{-\xi}$, has been demonstrated in the studies of the conventional MBL, induced by random~\cite{AQS4,MBL_num_dy3} or quasiperiodic~\cite{cold_atom3,MBL_num_a2} fields, for intermediate disorder strength. In this work, we reveal the power-law decay of imbalance in the ergodic side of the Stark MBL transition. By employing the ansatz  $\xi\propto\{|\gamma-\gamma_{c}|^{\nu},0\}$, which is widely used to estimate the critical disorder strength of the conventional MBL transition~\cite{cold_atom3}, we also extract the critical linear potential $\gamma_{c}$ from the data of imbalance. For the conventional MBL transition, it has been suggested that the $\gamma_{c}$ estimated by the imbalance is larger than that from the properties of highly-excited eigenstates~\cite{MBL_num_a2}. Our results indicate that the discrepancy between the $\gamma_{c}$ obtained from imbalance and the EE/PE of highly-excited eigenstates still exists for the Stark MBL transition. An illustration of the discrepancy remains an open question.

Recently, qualitative signatures of the Stark MBL, such as the memory of initial state with sufficiently large linear potential, have been observed in cold atoms~\cite{stark_exp1}, superconducting qubits~\cite{stark_exp2}, and trapped ions~\cite{stark_exp3}. We numerically demonstrate that the imbalance, as an experimentally feasible observable, can detect the transition point between ergodic and Stark many-body localized phase. Consequently, our work can increase the attention of further experimental studies of the quantum critical behaviors near the Stark MBL transition.

\begin{acknowledgments}
We acknowledge the discussion with Rubem Mondaini. This work was supported by National Science Foundation of China (NSFC) (Grant Nos. 11774406, 11934018),
Chinese Academy of Sciences (Grant No. XDB28000000), and Beijing Natural Science Foundation (Grant No. Z200009).

\end{acknowledgments}

\appendix

\section{ESTIMATION OF THE DENSITY OF STATES}

Different from the Heisenberg spin chain model, in which spectrum is symmetric and the maximum DoS locates in the middle of the spectrum($\epsilon^{*}=0.5$), the maximum DoS in the Bose-Hubbard model is much more dependent on on-site potential~\cite{MBL_num2,BH_MBL1}. Conventionally, the study of MBL pay attention to the eigenstates near the maximum DoS. Consequently, we should obtain the DoS as a function of normalized energy $\epsilon$, and ensure that the $\epsilon$ of highly excited eigenstates corresponds to the maximum DoS. Instead of numerically heavy ED, we employ the stochastic Chebyshev expansion method~\cite{Chebyshev,BH_MBL1}, which can efficiently produce the number of eigenvalues within any interval $[\epsilon_{i}, \epsilon_{i+1}]$, to accurately estimate the DoS.

\begin{figure}[]
	\centering
	\includegraphics[width=1\linewidth]{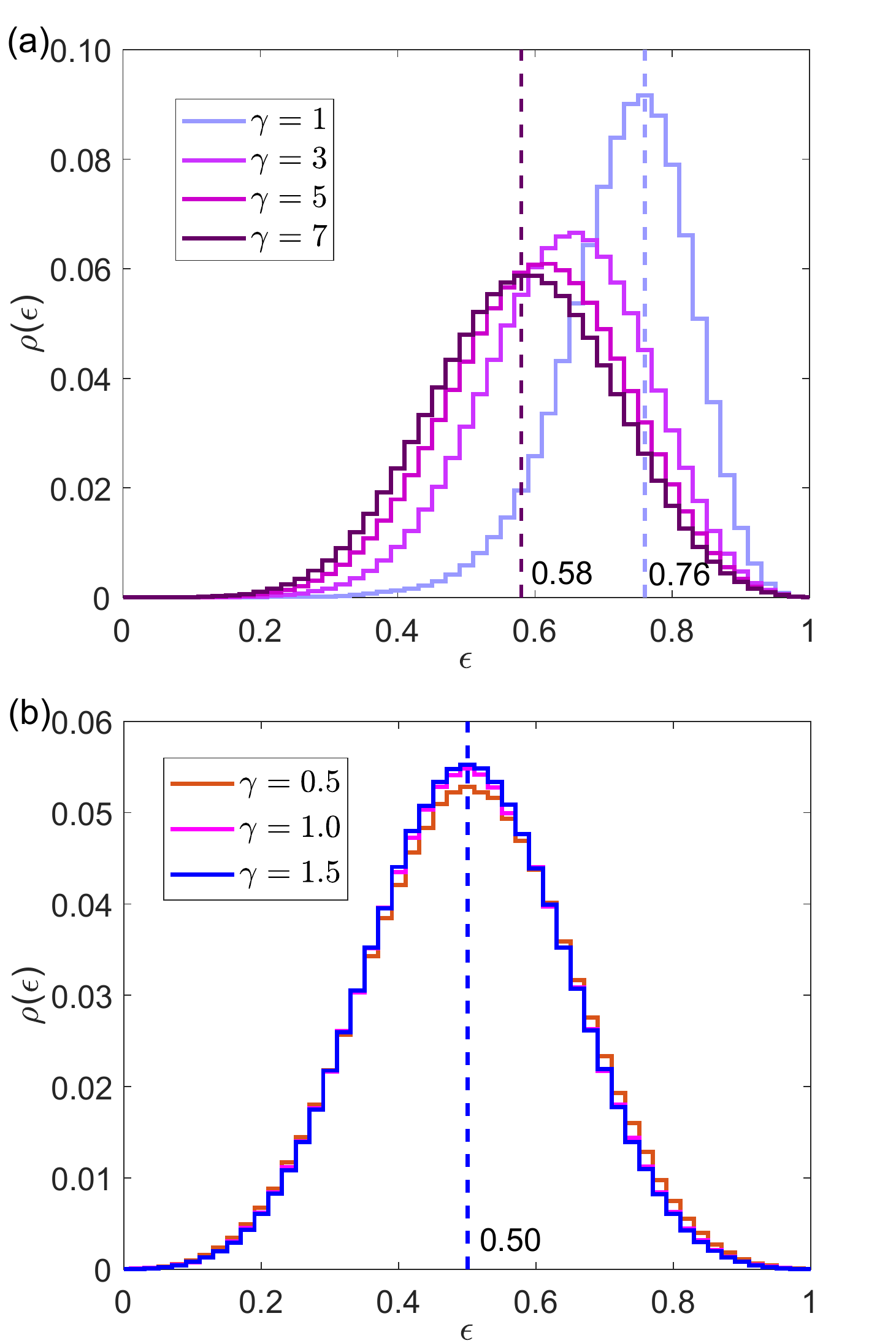}
	\caption{The DoS of different field strength $\gamma$ for (a) the Bose-Hubbard model with $\alpha=2,\ U=4,\ J=1$ and $L=14$, and (b) the all-to-all connected $XX$ model with $\alpha=2,\ g=0.5$ and $L=18$.}
	\label{fig:S1}
\end{figure}

Here, we display the DoS of the Bose-Hubbard model, i.e., the Hamiltonian (\ref{H_bh}), with several field strengths $\gamma$ in Fig.~\ref{fig:S1}(a), as well as the DoS of the all-to-all connected $XX$ model, i.e., the Hamiltonian (\ref{H_all_to_all}), in Fig.~\ref{fig:S1}(b) for comparison. It can be seen that as the field strength increases from $\gamma=1$ to $7$, the maximum DoS in the Bose-Hubbard model decreases from $\epsilon^{*}=0.74$ to $0.58$, while the maximum DoS in the all-to-all connected $XX$ model always locates near $\epsilon^{*}=0.5$.

Now, we can compute the eigenstates close to the maximum DoS, and investigate the properties of these eigenstates. In this work, we average the $\langle r\rangle$, $\overline{S^{EE}}$ and $\overline{S^{PE}}$ over 500 ($L = 10$) and 1000 ($L = 12,\ 14$) eigenstates closest to the maximum DoS $\epsilon^{*}$ for the Bose-Hubbard model (\ref{H_bh}), and over 200 ($L = 12$), 500 ($L = 14$) and 1000 ($L = 16,\ 18$) eigenstates for the $XX$ model (\ref{H_all_to_all}). Besides, to study the quench dynamics, we choose 50 initial states randomly within a narrow window near the maximum DoS $\epsilon^{*}$ ($|\epsilon-\epsilon^{*}|<0.05$ and $|\epsilon-\epsilon^{*}|<0.02$ for the Bose-Hubbard model and the $XX$ model, respectively).

\section{THE PAGE VALUE FOR THE BOSE-HUBBARD MODEL}

\begin{figure}[]
	\centering
	
	\includegraphics[width=1\linewidth]{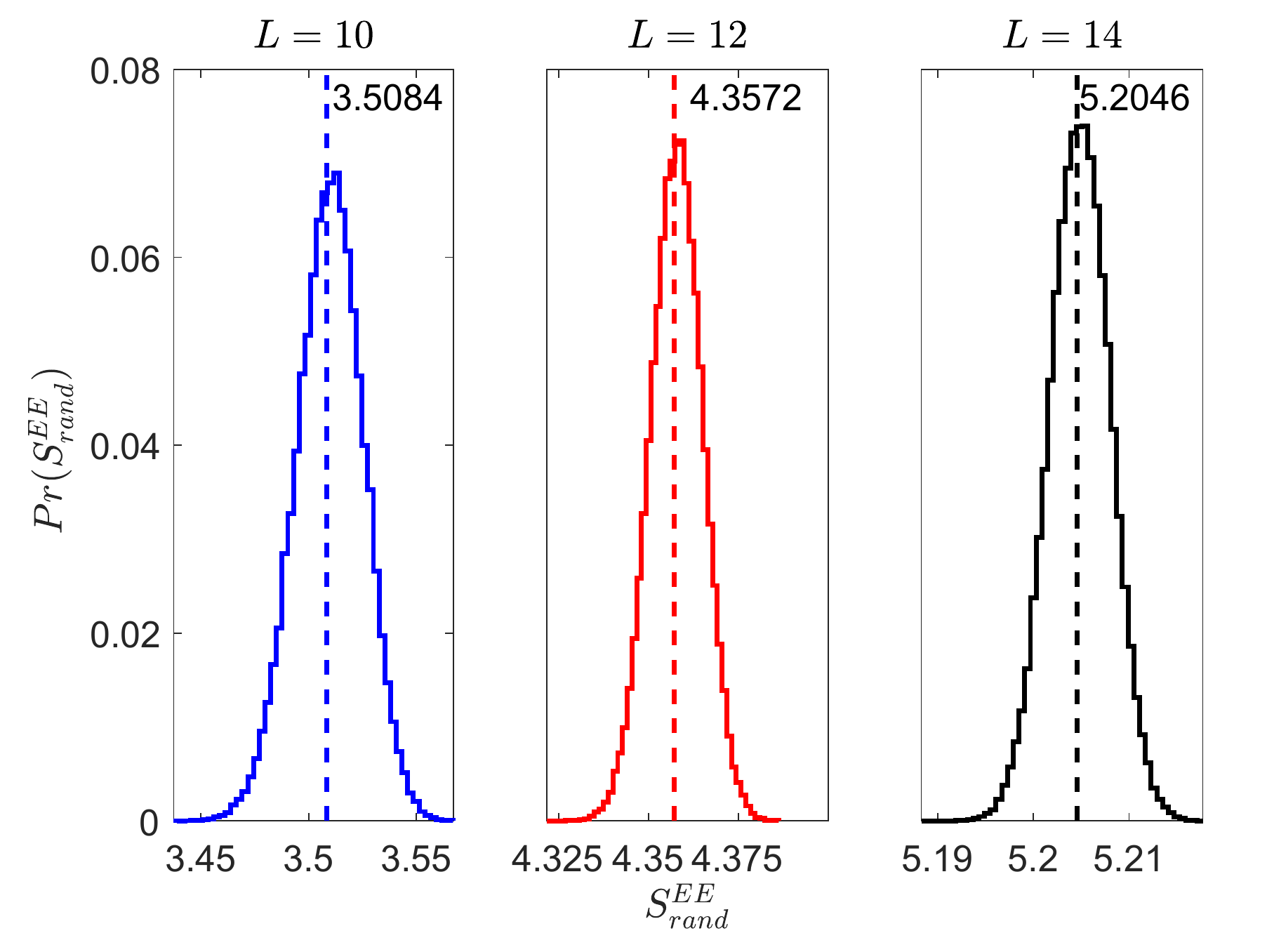}
	\caption{Histograms of the half-chain EE for 100,000 random pure states, calculated for several system sizes $L$. The dashed lines indicate the average values, which converge to the Page values.}
	\label{fig:S2}
	
\end{figure}

The Page value we focus on refers to the half-chain EE of a random pure state within the half-filling sector (the conserved total particle number $N=L/2$), which is expected to be the EE of an ergodic system at infinite temperature. For a spin-$\frac{1}{2}$ or fermion system with sizes $L$, the Page value satisfies $S_{T}^{EE}=0.5[L \ln (2)-1]$~\cite{Page}. The Page value for the Bose-Hubbard model, however, is not a simple generalization like $S_{T}^{EE}= \ln (\mathcal{M})-1/2$, with $\mathcal{M}$ being the dimension of half-chain subsystem Hilbert space, taking the conservation of total number of bosons into consideration. Here, we estimate the value directly by calculating the EE of random pure states within the half-filling sector. The distribution of the EEs for 100,000 random pure states is shown in Fig.~\ref{fig:S2}. The average EEs, i.e., the Page values $S_{T}^{EE}$, are $3.5084$, $4.3572$ and $5.2046$ for $L=10$, $12$ and $14$, respectively [see the dashed lines in Fig.~\ref{fig:S2}].

\section{AN ATTEMPT FOR USING THE BKT-TYPE SCALING TO DESCRIBE THE STARK MBL TRANSITION}

\begin{figure*}[]
    \centering
    \includegraphics[width=1\linewidth]{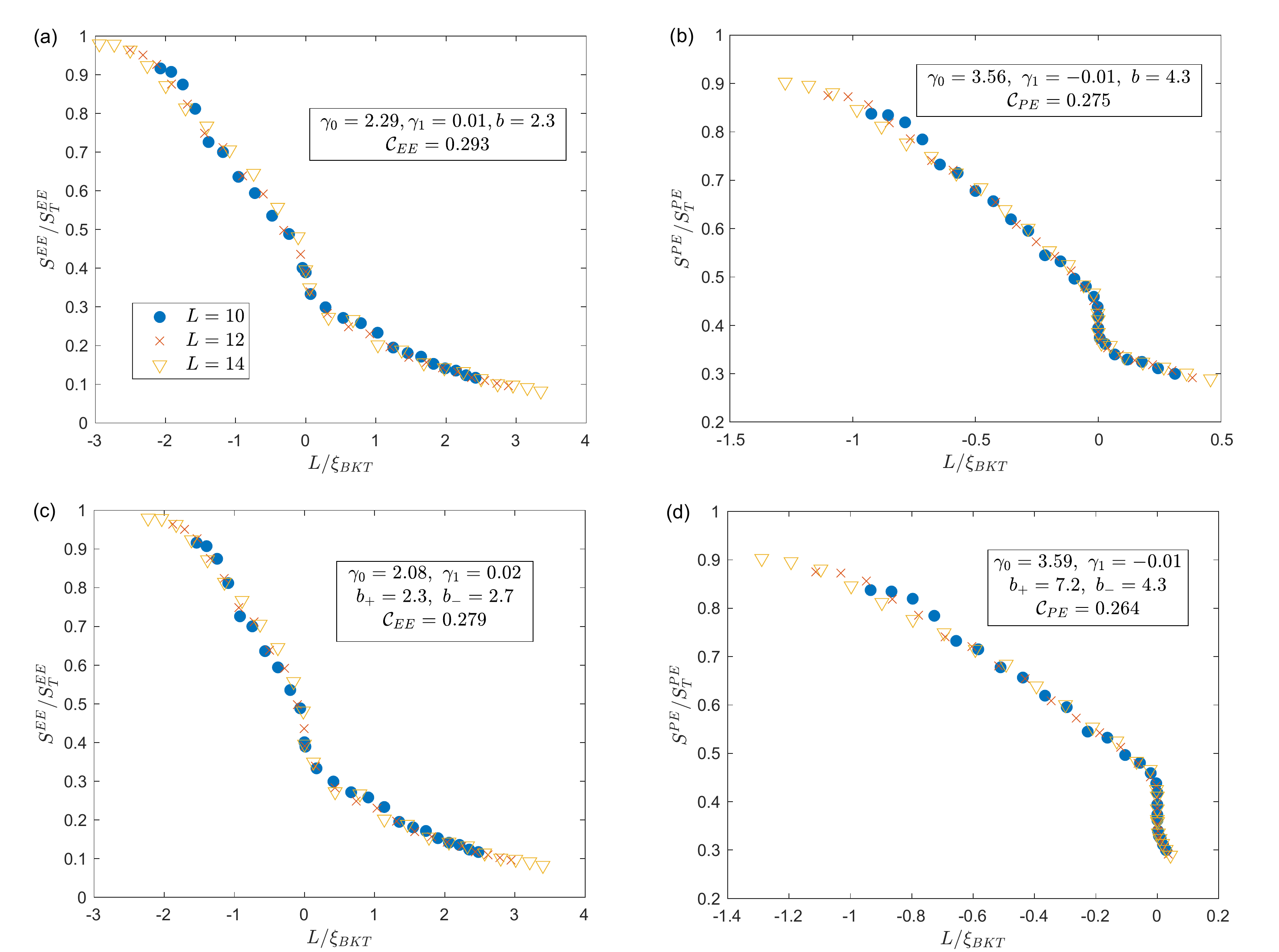}
    \caption{(a) The EE after the data collapse with a BKT-type correlation length. We use the crossing point ansatz $\gamma_c=\gamma_0+\gamma_1L$, assuming $b_{-} = b_{+} = b $. (b) is similar to (a) but for the PE. (c) is similar to (a) but with $b_{-} \neq b_{+} $. (d) is similar to (c) but for the PE.}
    \label{fig:BKT}
  \end{figure*}

Recently, several investigations~\cite{BKT1,BKT2,BKT3} have shown that the disorder-induced MBL transition follows
the BKT type of transition with a correlation length as
\begin{eqnarray}
\xi_{\mathrm{BKT}}=\exp \left\{\frac{b_{\pm}}{\sqrt{\left|W-W_{c}\right|}}\right\},
\label{BKT}
\end{eqnarray}
where $W_{c}$ is the critical disorder, and $b_{-}\ (b_{+})$ are parameters for $W < W_{c}\ (W > W_{c})$. The BKT-type scaling
seems to describe the disorder-induced MBL transition better than the power-law scaling $\xi_{0}\propto\{|W-W_{c}|^{\nu},0\}$~\cite{BKT2}.

Here, for the Stark MBL transition in the Bose-Hubbard model (\ref{H_bh}), we employ the BKT scenario, i.e., Eq.~(\ref{BKT}) with $W$ and $W_{c}$ replaced by $\gamma$ and $\gamma_{c}$, to perform the data collapse of the EE and PE shown in Fig.~\ref{fig3} (a) and (c), respectively. To find the best data collapse and quantitatively compare different scenarios, Ref.\cite{BKT2} introduces the cost function
\begin{eqnarray}
\mathcal{C}_{X}=\frac{\sum_{j=1}^{N_{\mathrm{p}}-1}\left|X_{j+1}-X_{j}\right|}{\max \left\{X_{j}\right\}-\min \left\{X_{j}\right\}}-1,
\label{cost_fun}
\end{eqnarray}
where $X_j$ denotes $\{S^{EE}/S^{EE}_{T}, S^{PE}/S^{PE}_{T}\}$ at different $\gamma$ and $L$, which is sorted according to nondecreasing values of $\operatorname{sgn}\left[\gamma-\gamma_c\right] L / \xi$. When we find the solutions of $b_{\pm}$ (or $\nu$) and $\gamma_c$ that minimize the cost function, the best data collapse is obtained.

First, we consider the BKT-type scaling, assuming $b_{-} = b_{+} = b$ and the crossing point $\gamma_c=\gamma_0+\gamma_1L$. The EE and PE after the data collapse are plotted in Fig.~\ref{fig:BKT}(a) and (b), respectively, with the minimum of cost function $\mathcal{C}_{EE}[\xi_{BKT}]\approx 0.293$ and $\mathcal{C}_{PE}[\xi_{BKT}]\approx 0.275$. For the sake of comparison, we also calculate the cost function with a power-law correlation length $\xi_0$ and a fixed crossing point $\gamma_c=\gamma_0$ (see Fig.~\ref{fig3}(b) and (d)), with the minimum $\mathcal{C}_{EE}[\xi_{0}]\approx 0.213$ and $\mathcal{C}_{PE}[\xi_{0}]\approx 0.264$.

We further loose the ansatz (\ref{BKT}) with $b_{-} \neq b_{+} $, and the results are shown in Fig.~\ref{fig:BKT}(c) and (d). In this case, the minimum of cost function is smaller than the previous ansatz with $b_{-} = b_{+} = b$ (see TABLE~\ref{T_cf}). It is seen that the BKT-type scaling doesn't seem to give a better description about the Stark MBL transition than the conventional framework of power-law divergence. However, we cannot exclude the BKT scenario, because the minimum of cost function of the BKT-type scaling is also relatively small.

  \begin{table}[htbp]
    \centering
    \begin{tabular}{|c|c|c|}
    \hline  $\mathcal{C}_{EE}\left[\xi_{0}\right]$ & $\mathcal{C}_{EE}^{b_+=b_-}\left[\xi_{\text {BKT }}\right]$& $\mathcal{C}_{EE}^{b_+ \neq b_-}\left[\xi_{\text {BKT }}\right]$\\
    \hline   0.213 & 0.293 &0.279\\
    \hline  $\mathcal{C}_{PE}\left[\xi_{0}\right]$ & $\mathcal{C}_{PE}^{b_+=b_-}\left[\xi_{\text {BKT }}\right]$& $\mathcal{C}_{PE}^{b_+ \neq b_-}\left[\xi_{\text {BKT }}\right]$\\
    \hline   0.264 & 0.275 &0.264\\
    \hline
    \end{tabular}
    \caption{The minimum of cost function of the BKT scaling and the power-law scaling for the EE and PE.}
    \label{T_cf}
    \end{table}

Whether the Stark MBL transition better agree with the BKT scenario remains an open question. One can explore this subject starting by studying the Stark MBL transition in a simple Heisenberg model~\cite{stark1,stark2}, where larger system size $L=24$ can be numerically achieved~\cite{MBL_num2}. Moreover, using the BKT scenario to explore the conventional MBL transition in the Bose-Hubbard model (\ref{H_bh}), which has been studied by employing the power-law scaling~\cite{BH_MBL1}, is a further investigation.

\section{THE EFFECT OF ANHARMONICITY ON DYNAMICAL BEHAVIOR IN THE BOSE-HUBBARD MODEL}

\begin{figure}[]
	\centering
	\includegraphics[width=1\linewidth]{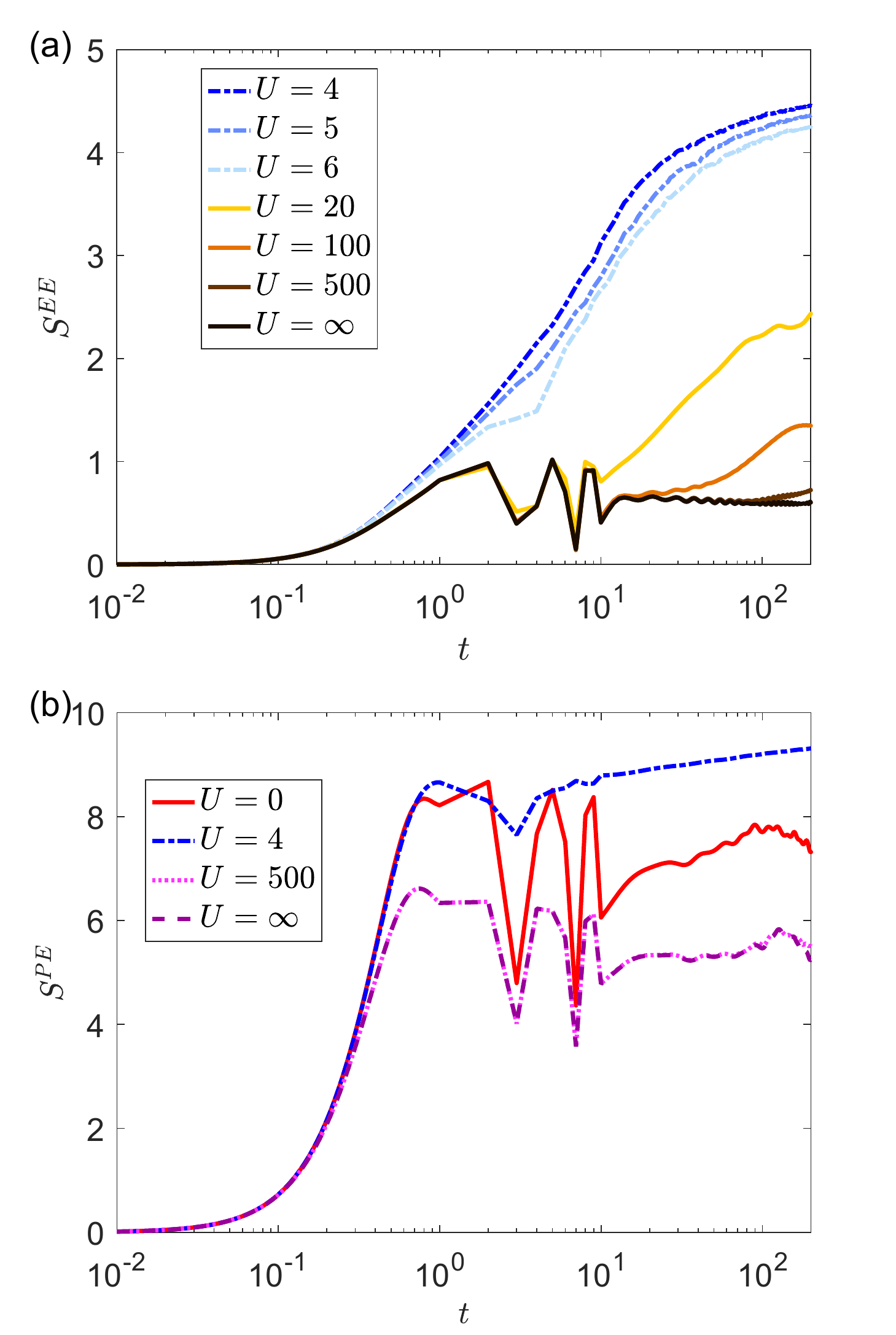}
	\caption{(a) The dynamics of EE in the Hamiltonian (1) with $L = 14,\ J = 1,\ \gamma = 2,\ \alpha=2$, and several values of anharmonicity $U$ for the initial charge-density wave state, i.e., $\left|\psi_{0}\right\rangle=\left|1010\ldots10\right\rangle$. The data of $U \geq 20$ have been smoothed by convolution. (b) is similar to (a) but for the dynamics of PE.}
	\label{fig:S3}
\end{figure}

In this appendix, we discuss the effect of anharmonicity $U$ on the quench dynamics in the Bose-Hubbard model. We initialize the system in a charge-density wave (CDW) state, i.e., $\left|\psi_{0}\right\rangle=\left|1010\ldots10\right\rangle$. The dynamical behavior of EE with different values of anharmonicity $U$ for the CDW state is shown in Fig.~\ref{fig:S3}(a). It can be seen that for the same field strength $\gamma=2$, the growth of EE slows down with the increase of the anharmonicity $U$. As the anharmonicity $U$ goes to infinity, the system degenerates to a chain of hard-core bosons, equivalent to the $XX$ model with the nearest-neighbor couplings. In this case, the on-site linear potential gives rise to the noninteracting Wannier-Stark localization, where the EE saturates quickly and the saturation value is suppressed significantly from the value in the interacting case. Moreover, as shown in Fig.~\ref{fig:S3}(b), the dynamical behavior of PE with $U=4$ is also distinguishable from the noninteracting bosonic case ($U=0$), and the hard-core limit ($U=500$ and $U=\infty$).

\begin{figure}[]
	\centering
	\includegraphics[width=1\linewidth]{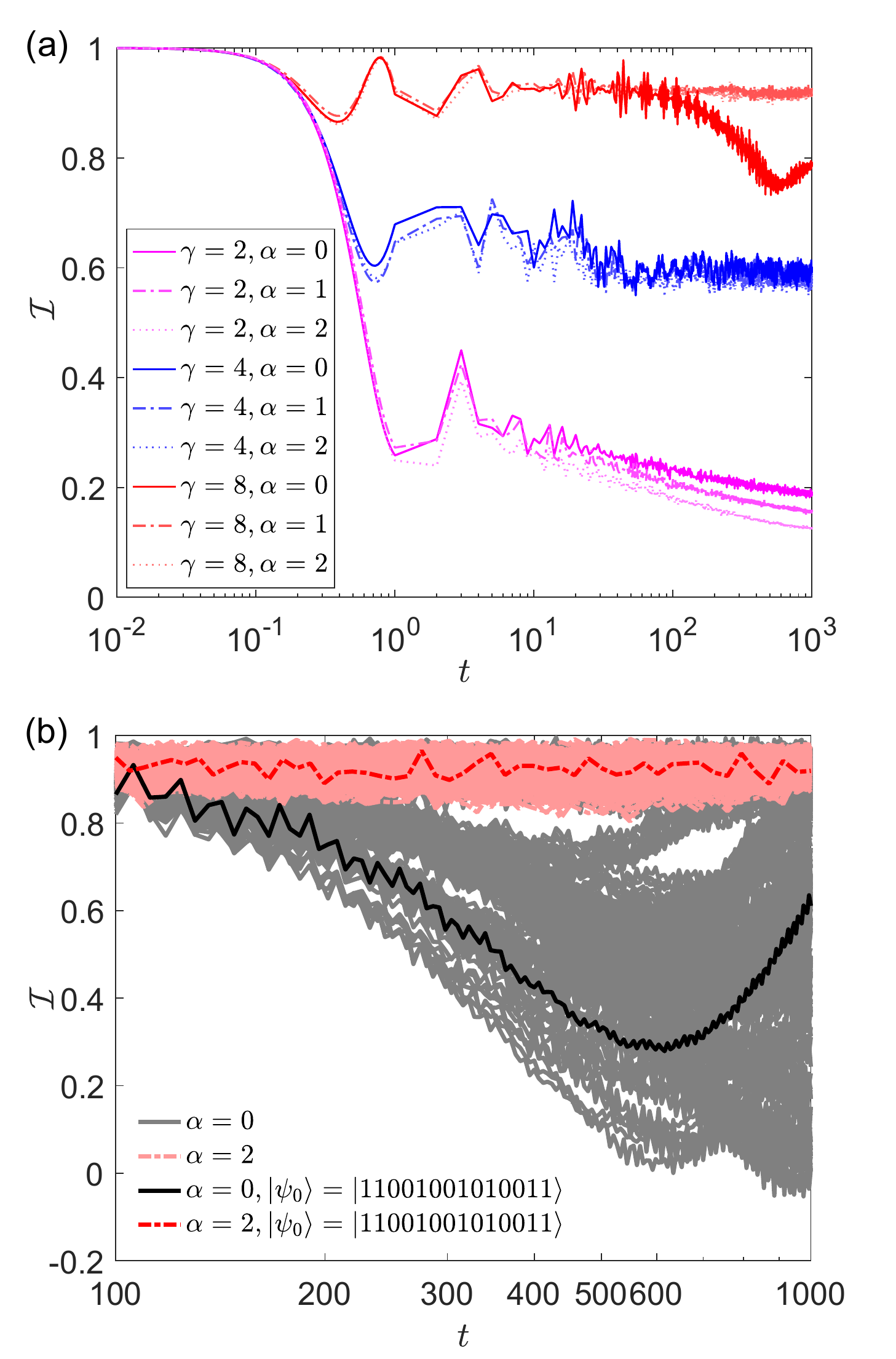}
	\caption{The dynamics of imbalance in the Hamiltonian (1) with $L=14$, $J=1$, $U=4$, and several field strengths $\gamma$ with different curvatures $\alpha$ on a logarithmic $x$ axis. (b) The dynamics of imbalance for 500 identical initial states with $\alpha=0$ and $\alpha=2$. The dynamics of $\left|\psi_{0}\right\rangle=\left|11001001010011\right\rangle$ is highlighted with black and deep red line for $\alpha=0$ and $\alpha=2$, respectively.}
	\label{fig:S4}
\end{figure}

 \section{THE EFFECT OF CURVATURE ON DYNAMICAL BEHAVIOR IN THE BOSE-HUBBARD MODEL}

 \begin{table*}[htbp]
 	
 	\centering
 	\begin{tabular}{|c|c|c|c|c|c|c|c|c|}
 		\hline
 		& & &\multicolumn{2}{c|}{}&\multicolumn{2}{c|}{}&\multicolumn{2}{c|}{} \\[-8pt]
 		\cline{4-9}
 		Curvature & Initial State $\left|\psi_{0}\right\rangle$ & $c_{\tilde{n}}$ & $a_1$ & $\left| i_{1}\right\rangle$ & $a_2$ & $\left| i_{2}\right\rangle$ & $a_3$ & $\left| i_{3}\right\rangle$ \\
 		& & & & & & & & \\[-10pt]
 		\hline
 		& & & & & & & & \\[-6pt]
 		$\alpha=0$ & $\left|11001001010011\right\rangle$ & -0.62 & -0.62 & $\left|11001001010011\right\rangle$ & -0.37 & $\left|1{\color{red}\textbf{0110}}001010011\right\rangle$ & -0.29 & $\left|110010010{\color{red}\textbf{0110}}1\right\rangle$ \\
 		\hline
 		& & & & & & & & \\[-6pt]
 		$\alpha=2$ & $\left|11001001010011\right\rangle$ & 0.82 & 0.82 & $\left|11001001010011\right\rangle$ & -0.29 & $\left|110010010100{\color{red}\textbf{20}}\right\rangle$ & -0.28 & $\left|{\color{red}\textbf{20}}001001010011\right\rangle$ \\
 		\hline
 	\end{tabular}
 	\caption{The three bases with the highest participation rate. The bases $\left| i_{1,2,3}\right\rangle$ correspond to coefficients $a_{1,2,3}$, with $a_{1,2,3}$ being the largest three modulus among all the basis participating in the eigenbasis $\left| E_{\tilde{n}}\right\rangle$, when initial state $\left|\psi_{0}\right\rangle=\left|11001001010011\right\rangle$ for different curvature $\alpha=0$ and $\alpha=2$.}
 	\label{table:T2}
 	
 \end{table*}

 To verify the effect of curvature $\alpha$ in Eq.~(\ref{linear}) on the dynamical behaviors in the Bose-Hubbard model, we calculate the imbalance for different curvatures $\alpha$ [see Fig.~\ref{fig:S4}(a)]. For small and intermediate field strengths $\gamma$, the curvature influences the decay exponent $\xi$ of imbalance. When the field strength $\gamma$ just barely drives the system into a localized regime, where decay exponent is already indistinguishable from zero, the curvature does not show a significant effect [see the results of $\gamma=4$, i.e., the blue lines in Fig.~\ref{fig:S4}(a)]. As the field strength increases further, the dipole moment is becoming conserved approximately, and Hilbert space begins to fragment or shatter. In this regime, the presence of curvature is the key to hold back the fragmentation and maintain the memory of the initial state [see the results of $\gamma=8$, i.e., the red lines in Fig.~\ref{fig:S4}(a)].

 We further check the result of $\gamma=8$ with ED for 500 identical initial states for both $\alpha=0$ and $\alpha=2$. The normalized energies of the initial states belong to $[\epsilon^{*}-0.02, \epsilon^{*}+0.02 ]$. The maximum DoS is almost the same in both cases, which is $\epsilon^{*} \approx 0.57$. As seen in Fig.~\ref{fig:S4}(b), for $\alpha=0$, the dynamics depends strongly on the initial states, and many of them lose the memory of initial states during the evolution even with $\gamma=8$. However, for $\alpha=2$, the dynamics of imbalance indicates that the memory of all different initial states is retained.

 Specifically, we highlight the dynamics of the identical initial state $\left|\psi_{0}\right\rangle=\left|11001001010011\right\rangle$ with the black and deep red line for $\alpha=0$ and $\alpha=2$, respectively, which exhibit dramatically different dynamical behaviors. To better understand the different behaviors, we expand the initial state in terms of the eigenstates of the Hamiltonian $\ket{E_n}$, i.e., $\left|\psi_{0}\right\rangle=\sum_{n}c_n\left|E_n\right\rangle$ such that
 \begin{equation}
 \left|\psi_{t}\right\rangle=e^{-i\hat{H}t}\left|\psi_{0}\right\rangle=\sum_{n}e^{-iE_{n}t}c_n\left|E_n\right\rangle.
 \end{equation}
 Then, we label the eigenstate with the maximum modulus of the expansion coefficient as $\ket{E_{\tilde{n}}}$, and the corresponding expansion coefficient $c_{\tilde{n}}=\max(\{c_n\})$. In other words, the eigenstate $\ket{E_{\tilde{n}}}$ dominantly participate in the dynamical behavior. We can express the eigenstate $\ket{E_{\tilde{n}}}$ in the Fock basis
 \begin{equation}
 \left|E_{\tilde{n}}\right\rangle=a_1\left|i_1\right\rangle+a_2\left|i_2\right\rangle+a_3\left|i_3\right\rangle+\ldots+a_{\mathcal{N}}\left|i_{\mathcal{N}}\right\rangle ,
 \end{equation}
 where $|i_k\rangle \in \{|n_{1}n_{2}...n_{L}\rangle| \sum_{j=1}^{L} n_{j} = \frac{L}{2}, n_{j}\in \mathbb{N}, 0\leq n_{j} \leq \frac{L}{2}\}$, $\mathcal{N}$ is the dimension of Hilbert space, and $a_k=\left\langle i_{k}\mid E_{\tilde{n}}\right\rangle$ is sorted by modulus in descending order.

 As can be seen in TABLE~\ref{table:T2}, $\left|i_1\right\rangle$ is the same with initial state $\left|\psi_{0}\right\rangle$ in either case, and thus $|c_{\tilde{n}}|=|a_1|$. But there is a dramatic difference in $\left| i_{2,3}\right\rangle$  between $\alpha=0$ and $\alpha=2$. For pure linear potential, the dynamics is dominated by the movement of ``dipoles''. However, if the curvature is present, eigenstates are close to product states of localized single particle, similar to the picture of local integrals of motion (LIOM) in conventional MBL~\cite{stark5}.

 It is worth noting that for $\alpha=0$, there are still a number of initial states exhibiting little loss of initial-state memory, most of which have ``difficulty'' in forming a dipole, including but not limited to low density of domain walls~\cite{Doggen2021}.

 \section{MORE RESULTS OF THE $XX$ MODEL WITH ALL-TO-ALL CONNECTIVITY}

 \begin{figure}[]
 	\centering
 	\includegraphics[width=1\linewidth]{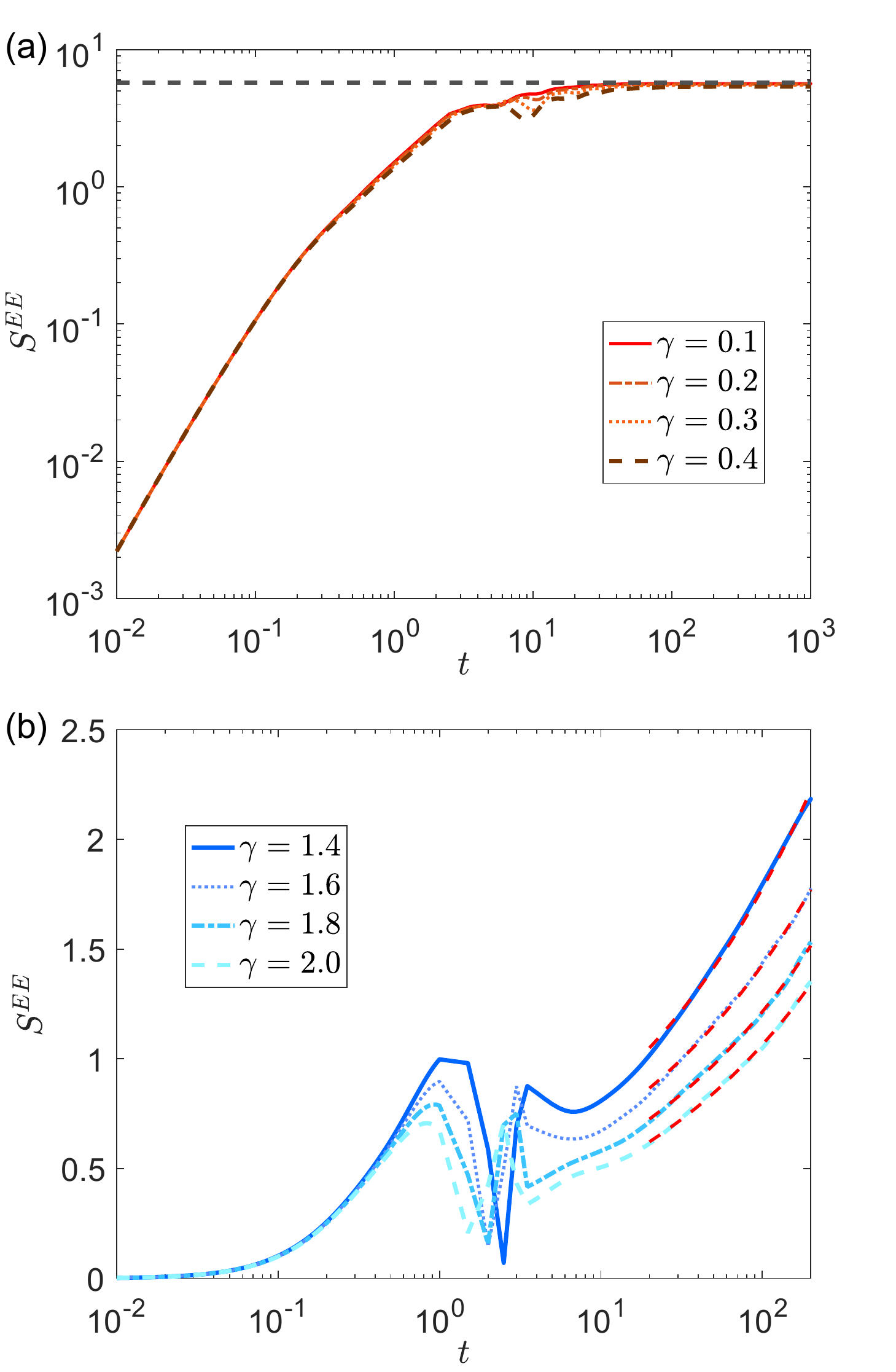}
 	\caption{(a) The dynamics of EE in the Hamiltonian (2) with $L = 18,\ g = 0.5,\ \alpha = 2$ and several field strengths $\gamma$ smaller than the critical value $\gamma_c$. The horizontal dashed line marks the Page value. (b) The dynamics of EE in the same system in (a) but for larger $\gamma$. The red dashed lines correspond to the fits to the power-law form $S(t)\propto t^{\beta}$, and $\beta \approx 0.31,\ 0.32,\ 0.32,\ 0.33$ for $\gamma= 1.4,\ 1.6,\ 1.8,\ 2$, respectively.}
 	\label{fig:S5}
 \end{figure}

 \begin{figure}[]
 	\centering
 	\includegraphics[width=1\linewidth]{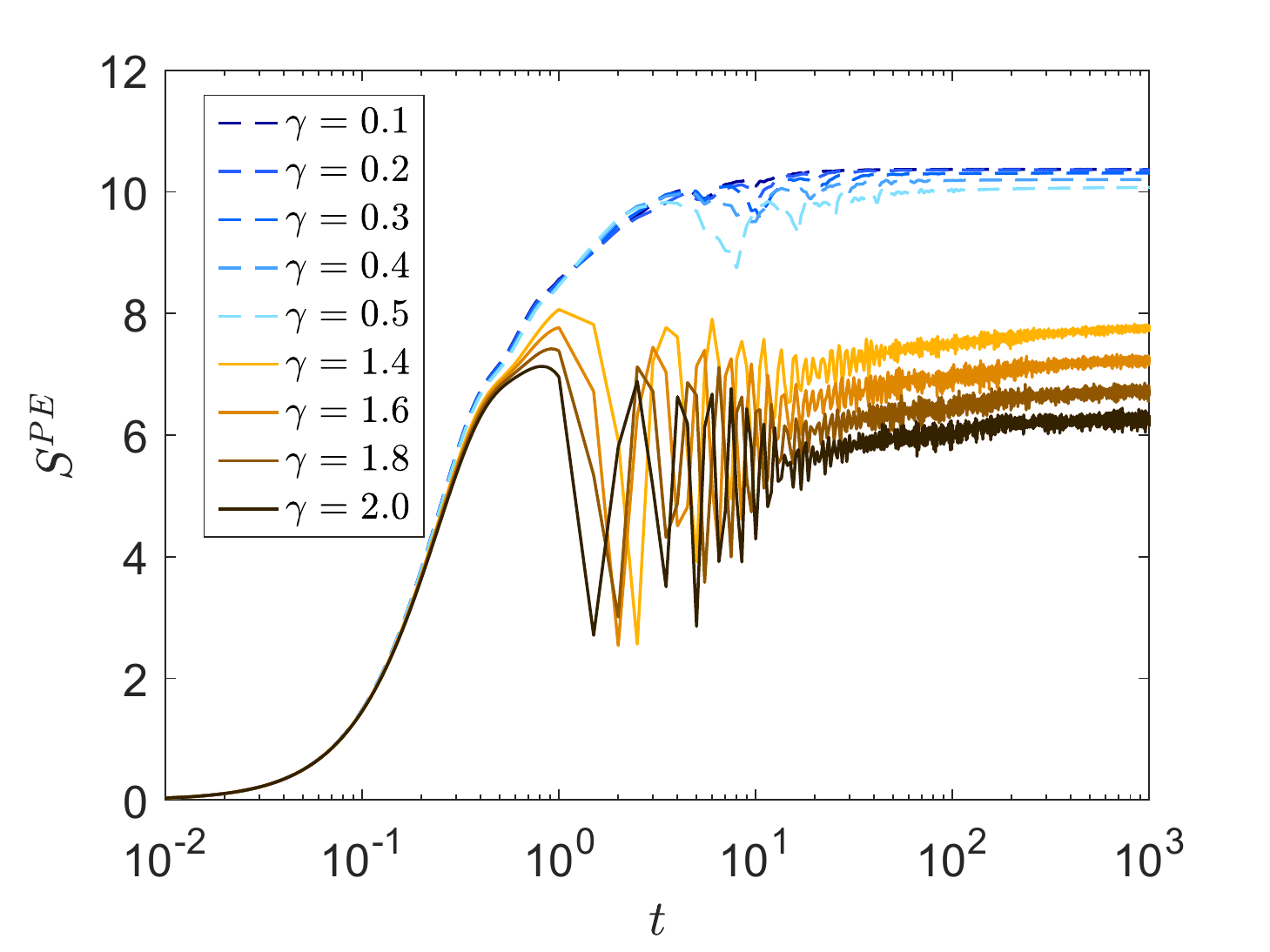}
 	\caption{The dynamics of PE in the Hamiltonian (2) with $L = 18,\ g = 0.5,\ \alpha = 2$ and several field strengths $\gamma$.}
 	\label{fig:S6}
 \end{figure}

In this appendix, we present the time evolution of the EE and PE in the $XX$ model with all-to-all connectivity, i.e., the Hamiltonian (\ref{H_all_to_all}).

In Fig.~\ref{fig:S5}, we show the time evolution of EE $S^{EE}(t)$ with several field strengths. For $\gamma<\gamma_c$ [see Fig.~\ref{fig:S5}(a)], the EE approaches the Page value quickly with a ballistic spreading. With increasing $\gamma$ [see Fig.~\ref{fig:S5}(b)], the spreading of entanglement slows down. Note that although the EE grows much slower than that in ergodic cases, it exhibits a power-law growth $S(t)\propto t^{\beta}$ ($\beta$ is about 0.32) at a long time, rather than a logarithmic growth. Similar behaviors of power-law growth of EE in the MBL phase have been observed in the $XY$ model and the Heisenberg model with power-law interactions~\cite{Safavi-Naini2019}. Thus, we can attribute the power-law growth of EE to the non-local interaction, for which further investigation is required.

We also study the dynamics of PE, and the results are plotted in Fig.~\ref{fig:S6}. It can be seen that the growth of PE with increasing $\gamma$ slows down, which exhibits a similar behavior of the time evolution of the PE in the Bose-Hubbard model, i.e., the results in Fig.~\ref{fig5}.

\bibliography{reference_stark}

\end{document}